\documentclass[english,usenatbib]{mn2e}
\usepackage{amsmath}
\usepackage{color}
\usepackage{babel}
\usepackage{array}
\usepackage{refstyle}
\usepackage{booktabs}
\usepackage{units}
\usepackage{multirow}
\usepackage{graphicx}
\usepackage[unicode=true,pdfusetitle,
 bookmarks=true,bookmarksnumbered=false,bookmarksopen=false,
 breaklinks=false,pdfborder={0 0 1},backref=section,colorlinks=false]
 {hyperref}

\makeatletter
\providecommand{\tabularnewline}{\\}

\usepackage{aas_macros}
\usepackage{epstopdf}
\PrependGraphicsExtensions{.svg}
\makeatother

\begin{document}

\title[Growth and Anisotropy of Ionization Fronts]{
    Growth and anisotropy of ionization fronts near high
    redshift quasars in the MassiveBlack simulation}

\author[Yu Feng et al.]{
    Yu Feng$^{^{1}}$\thanks{yfeng1@cmu.edu}, 
    Rupert A.C. Croft$^{^{1,2}}$,
    Tiziana Di Matteo$^{^{1,2}}$, and 
    Nishikanta Khandai$^{^{1}}$ \\
    $^{^{1}}$ McWilliams Center for Cosmology, 
    Department of Physics, 
    Carnegie Mellon University,
    Pittsburgh, PA 150213, USA\\
    $^{^{2}}$ Physics Department, University of Oxford,
    Keble Road, Oxford OX1 3RH, UK\\
  }
\maketitle
\begin{abstract}
We use radiative transfer to study the growth of ionized
    regions around the brightest, $z=8$ quasars in a large
    cosmological hydrodynamic simulation that includes black
    hole growth and feedback (the MassiveBlack simulation).
    We find that in the presence of the quasars the
    comoving HII bubble radii reach $\unit[10]{Mpc/h}$ after
    20 Myear while with the stellar component alone the HII
    bubbles are smaller by at least an order of magnitude.
    Our calculations show that several features are not
    captured within an analytic growth model of Stromgren
    spheres.  The X-ray photons from hard quasar spectra
    drive a smooth transition from fully neutral to
    partially neutral in the ionization front. However the
    transition from partially neutral to fully ionized is
    significantly more complex. We measure the distance to
    the edge of bubbles as a function of angle and use the
    standard deviation of these distances as a diagnostic of
    the anisotropy of ionized regions. We find that the
    overlapping of nearby ionized regions from clustered
    halos not only increases the anisotropy, but also is the
    main mechanism which allows the outer radius to grow. We
    therefore predict that quasar ionized bubbles at this
    early stage in the reionization process should be both
    significantly larger and more irregularly shaped than
    bubbles around star forming galaxies. Before the star
    formation rate increases and the Universe fully
    reionizes, quasar bubbles will form the most striking
    and recognizable features in 21cm maps.
\end{abstract}

\section{Introduction}
The current consensus is that the contribution to the
    global ionizing budget from quasars during the Epoch of
    Reionization (EoR) is small compared to that from early
    stars and galaxies \citep[see, eg,
    ][]{2009JCAP...03..022L,1996ApJS..102..191G,2011ASL.....4..228T}.
    The EoR began as the population III stars and
    primordial galaxies ionized their most immediate
    vicinity, as studied by \citet{2008MNRAS.384.1080T,
    2008ApJ...684...18C, 2011MNRAS.417.2264V}.
    On the other hand, the characteristic proper radius of
    ionizing bubbles at the end of EoR is constrained to be
    on the order of $\unit[10]{Mpc/h}$
    \citep{2007MNRAS.376L..34M, 2007MNRAS.380L..30A,
    2004Natur.432..194W}, and the quasar contribution is
    limited to be less than 14\% \citep{2007MNRAS.374..627S}
    of the total.  Even though the contribution to global
    reionization by quasars is constrained in this way,
    bright quasars may still leave a signature on the growth
    of individual ionized regions.  Several authors have
    investigated this signature in mock 21cm emission
    spectra taken from simulations of isolated quasars
    \citep{2012MNRAS.424..762D, 2011arXiv1111.6354M,
    2010ApJ...723L..17A, 2008MNRAS.391.1900D,
    2008MNRAS.386.1683G} at redshift $z \sim 8$.  In
    observations, an object recently reported by
    \citet{2011Natur.474..616M}, ULAS J1120+0641, has a
    luminosity of $6.3\times10^{13}L_{\odot}$ at $z \sim 7$
    and a proper near-zone radius of less than
    $\unit[2]{Mpc/h}$.  The near-zone radius is consistent
    with the possibility of bright quasar driven growth in a
    near neutral intergalactic medium background \citep{2011MNRAS.416L..70B}.

In this paper we study the growth of the ionization front
    of bright quasars in an almost neutral cosmological
    context.  The quasars and their surrounding medium are
    selected from a large hydrodynamic simulation \citep[the
    MassiveBlack simulation, introduced in][]{2012ApJ...745L..29D}, 
    and then post-processed with a radiative transfer code.
    This allows us to simulate 8 rare quasars using
    reasonable computing resources.  Our focus is on the
    evolution and properties of the largest individual
    ionized bubbles, the sources that produce them, and the
    relationship between the two.  Because the simulation
    forms quasars and star forming galaxies ab initio, we
    are able to make use of the luminosity and positions of
    the radiation sources that the simulation produces,
    rather than setting them in by hand. However we do not
    deal with the full reionization of the volume of the
    simulation, which would require following the evolution
    of the entire density and ionization field from high
    redshifts down to at least $z=6$.  Instead we restrict
    ourselves to the growth of ionized regions around an
    early period in this process (at $z=8$), where the
    photon path lengths are still smaller than the
    computational sub-volumes we analyze. We leave the study
    of the full reionization of the volume to future work.

\section{Hydrodynamic simulation and density field}

The SPH output we use is from the MassiveBlack simulation
    \citep[see][for further details]{2012ApJ...745L..29D, 2012MNRAS.424.1892D,
    2012MNRAS.423.2397K}, which was run
    with a $\Lambda\rm CDM$ cosmology with parameters
    $(\Omega_{\Lambda}, \Omega_{M}, \Omega_{b}, h,
    \sigma_{8}) = (0.74, 0.26, 0.044, 0.72, 0.8)$.  A total
    number of $2\times3200^{3}$ gas and dark matter
    particles were followed in a box of
    $\unit[0.75]{Gpc^{3}}$ from redshift $z=159$ to
    redshift $z=4.75$.  This simulation is by far the
    largest cosmological hydrodynamics simulation run with
    the {\small P-GADGET} program. This run not only contains
    gravity and hydrodynamics but also the extra physics
    (sub-grid modeling) for star formation
    \citep{2003MNRAS.339..289S}, black holes and associated
    feedback processes.

The basics aspects of the black hole accretion and
    feedback model \citep{2008ApJ...676...33D} consist of
    representing black holes by collisionless particles that
    grow in mass (from an initial seed black hole) by
    accretion of gas in their environments. The accretion rate
    is given by 
\begin{equation}
  \dot{M} = \min (M_{\rm Bondi}, M_{\rm Edd})\,,
\end{equation}
    and
\begin{equation}
  \dot{M}_{\rm Bondi} = \frac 
      {4 \pi G^2 M_{BH}^2 \rho}{(c_s^2 + v^2)^{3/2}},
\end{equation}
    where $\rho$ and $c_s$ are the density and sound speed
    of the ISM gas respectively, and $v$ is the velocity of
    the black hole relative to the gas, and $\dot{M}_{\rm
    Edd} = L_{\rm Edd} / (\eta c^2)$, where $L_{\rm Edd} =
    \unit[1.26 \times 10^{38}]{erg s^{-1}}$,
    is the Eddington Luminosity. At the high
    redshift and high halo masses we are carrying out the
    analysis here, black holes are growing at their
    Eddington rates \citep{2012MNRAS.424.1892D} making any
    detail of the sub-grid models for black holes virtually
    irrelevant.  

For the black hole feedback, a fraction of the radiative
    energy released by the accretion of material is assumed to
    couple thermally to nearby gas and influence its motion
    and thermodynamic state.  The radiated luminosity,
    $L_{\rm r}$, from the black hole is related to accretion
    rate, $\dot {M}_{\rm BH}$, as
\begin{equation}
  \label{eq:bh-luminosity}
  {{L_{\rm bol}}= \eta {\dot {M}_{\rm BH} c^2}}, 
\end{equation}
    where we take the standard mean value $\eta =0.1$.  Some
    coupling between the liberated luminosity and the
    surrounding gas is expected: in the simulation 5\% of
    the luminosity is isotropically deposited as thermal
    energy in the local black hole kernel, providing some
    form of feedback energy.  
    
With our simulations the activity of quasars is directly
    derived from the accretion history of rapidly growing
    super-massive black holes, and their fueling is driven
    from the large scales and occurs through high density
    cold flows along the cosmic filaments
    \citep{2012ApJ...745L..29D}.  

The luminosity of the stars and galaxies can be derived
    from the star formation rate history which is provided
    by the multiphase star formation model in the simulation
    that depends on a single free parameter, $t_\star$ the
    global star formation timescale
    \citep{2003MNRAS.339..289S}. The multiphase star
    formation model also provides a mechanism to remove the
    self-shielded interstellar medium (ISM) from the matter
    density field, leaving only the intergalactic medium (IGM)
    density for the radiative transfer simulation.

In the multiphase star formation model, a dense gas
    particle is divided into a non-star forming IGM
    component and a star forming ISM component: 
\begin{equation}
  m = \begin{cases}
    m_{\rm IGM} + m_{\rm ISM} = 
    (1-x) m + x m, & \mbox{if } \rho > \rho_{\rm th}, \\
    m_{\rm IGM}, & \mbox{if } \rho \le \rho_{\rm th}, \\
  \end{cases}
\end{equation}
    where $x$ is the mass fraction of the ISM component. The
    threshold density $\rho_{\rm th}$ is determined from the
    global star formation time scale $t_\star$, as described
    in \citet{2003MNRAS.339..289S}. The IGM component of gas
    particles forms a hot ambient medium and occupies the
    entire volume of the particle. The ISM component of gas
    particles condenses into cold star-forming clouds which
    are self-shielded from cosmic radiative transfer, except
    that they may host stellar radiative sources. As a
    result we excise them from the matter density field when
    performing the ray tracing calculation. By doing this we
    also assume that their small cross-section would not
    have affected ray tracing for the rest of the gas.  The
    ISM fraction, $x$, is an increasing function of the mean
    density of the gas particle, effectively removing the
    densest particles from the density field in a manner
    similar to the threshold method used to calculate the
    clumping factor by \citet{2009MNRAS.394.1812P}, and, the
    removal of the cold high density gas in the X-ray
    emission calculation by \citet{2001ApJ...557...67C}. 

We also note that in MassiveBlack the mean baryon density
    (IGM + ISM) around quasar sources can be as high as
    $\unit[60]{cm^{-3}}$. Were the ISM not excised from the
    ray tracing, the high mean density would shield off the
    radiation and prevent the growth of any cosmic scale
    ionized regions. 

\section{Selection of Quasars}

We use the quasars and density field of the $z=8$ snapshot
   of MassiveBlack in this study. There are two reasons for
   this choice:
\begin{itemize}
  \item
    The EoR in MassiveBlack is modeled by a uniform UV
    background radiation field that is introduced near the
    end of the EoR ($z=6$) in the optically thin
    approximation \citep[see e.g.][]{2007MNRAS.382..325B}.
    By choosing an earlier redshift we do not contaminate
    the ionization fronts with this global radiation field.
  \item
    Extremely bright quasar sources at high redshifts are
    rare objects. Limited by the $\unit[0.75]{Gpc^3}$ volume
    of MassiveBlack, we cannot find many bright quasar
    systems at a much higher redshift than $z=8$.
\end{itemize} 

We select a quasar system based on halo mass, taking the 10
    most massive halos, numbered from 0 to 9.  We note that
    in general larger halos host brighter quasars unless (i)
    the quasar system is turned off by feedback, or (ii) the
    quasar system has not yet grown its black hole mass
    significantly.  8 unique hosting sub-volumes
    ($\unit[50]{Mpc/h}$ per side) are identified, which we
    refer as sub-volume 0 to 7 throughout the rest of the
    paper: Three of the halos (halo number 0, 3 and 5 in
    Table \ref{tab:Total-ionization-photons}) are located at
    spatial locations within sub-volume 0. 

\begin{table}
  \centering%
  \begin{tabular}{cccccccc}
  \toprule 
  \multirow{2}{*}{\#H} & 
  \multirow{2}{*}{\#V} & 
  \multirow{2}{*}{$M_{\rm HALO}$} & 
  \multirow{2}{*}{$M_{\rm BH}$} &
  \multicolumn{2}{c}{Total} & 
  \multicolumn{2}{c}{Center} \tabularnewline
  \cmidrule{5-8} 
  & & &  & QSO & STR & QSO & STR\tabularnewline
  \midrule
  \midrule 
  0 & 0 & 148.35 & 74.6 & 8.35 & 3.38 & 7.24 & 0.94\tabularnewline
  \midrule 
  1 & 1 & 77.9 & 49.6 & 4.97 & 3.45 & 4.43 & 0.79\tabularnewline
  \midrule 
  2 & 2 & 75.5 & 34.7 & 2.06 & 3.93 & 1.39 & 0.45\tabularnewline
  \midrule 
  3 & 0 & 84.9 & 7.3 & 8.35 & 3.71 & 0.49 & 0.22\tabularnewline
  \midrule 
  4 & 3 & 60.1 & 119.0 & 8.18 & 2.72 & 7.76 & 0.54\tabularnewline
  \midrule 
  5 & 0 & 67.1 & 16.0 & 8.35 & 3.71 & 0.56 & 0.21\tabularnewline
  \midrule 
  6 & 4 & 70.8 & 5.5 & 0.50 & 2.72 & 0.06 & 0.16\tabularnewline
  \midrule 
  7 & 5 & 65.6 & 13.2 & 0.95 & 1.98 & 0.69 & 0.30\tabularnewline
  \midrule 
  8 & 6 & 66.6 & 6.4 & 1.29 & 3.82 & 0.20 & 0.21\tabularnewline
  \midrule 
  9 & 7 & 64.9 & 7.2 & 0.38 & 2.22 & 0.15 & 0.15\tabularnewline
  \bottomrule
  \end{tabular}

  \caption{UV Flux of Sub-volumes. 
    The first column \#H (0 - 9)
    identifies the ten most massive halos.  The second
    column \#V (0 - 7) identifies the eight unique
    sub-volumes that contain the halos.  The halo mass
    ($M_{\rm HALO}$) is in units of
    $\unit[10^{10}]{M_{\odot}h^{-1}}$.  The blackhole mass
    ($M_{\rm BH}$) is in units of
    $\unit[10^{6}]{M_{\odot}h^{-1}}$.  The flux is
    in units of $\unit[10^{55}]{sec^{-1}}$.  Halos 0, 3, and
    5 are clustered and belong to the same sub-volume 0.
    See Appendix \ref{sub:Luminosity-Model} for the definition of the spectra.  }
  \label{tab:Total-ionization-photons}
\end{table}
\begin{figure*}
  \includegraphics[width=0.70\textwidth]{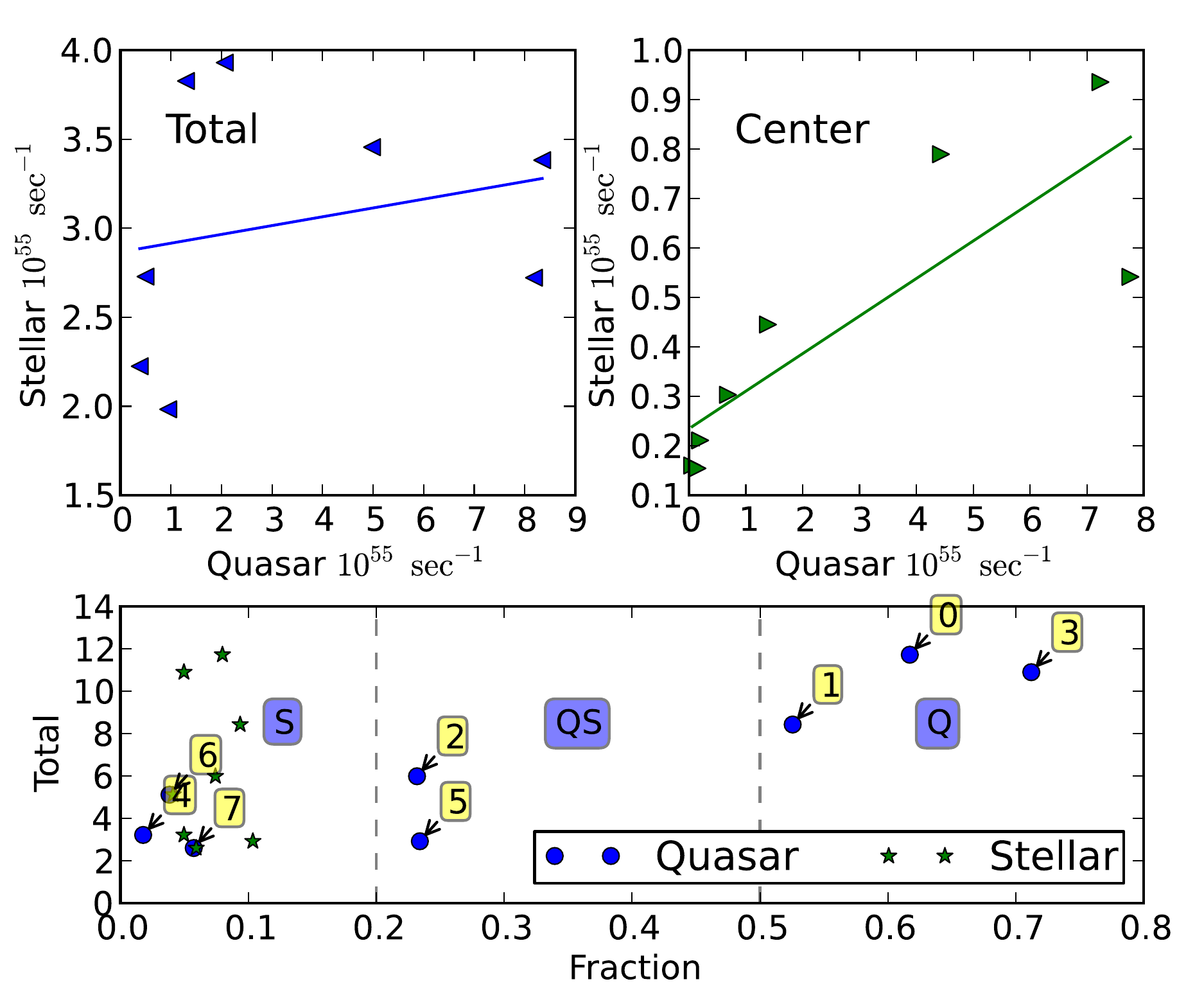}
  \caption{Central and total ionizing photon flux. 
    The left panel compares the total
    UV photon flux of the stars and the quasars
    within the sub-volumes. The right panel compares the
    UV photon flux from the sources within
    $\unit[1]{Mpc/h}$ of the center.  The lines
    represent the linear regression of the data. In the
    bottom panel the central photon flux of the quasar sources
    and the central photon flux of the stellar sources as a
    fraction of the total flux are shown.  It is
    interesting to observe that the central stellar photon flux is
    always a small fraction of the total flux yet
    the central quasar can contribute a significant fraction
    of the total. See Appendix \ref{sub:Luminosity-Model}
    for the definition of the spectra.
    }
  \label{fig:Sample-Luminosity}
\end{figure*}

Next, we compute the ionizing photon flux due to the quasars and
    star formation occurring in each of the sub-volumes. The
    details of the assignment are described in Appendix
    \ref{sub:Luminosity-Model}. In addition to the full
    volume flux, we also define a Central Source Flux, which
    is the flux of sources within the $\unit[1]{Mpc/h}$
    radius of the center of the sub-volume. The sources in
    the center are associated with the ionized bubble of the
    central halo. The ionizing photon flux of the sub-volumes
    is summarized in Table
    \ref{tab:Total-ionization-photons} and Figure
    \ref{fig:Sample-Luminosity}.

There is an expected correlation between the central
    quasar and central stellar flux. Based on the ratio of
    the central flux to total ionizing flux within the
    sub-volume, as well as the composition (quasar or
    stellar) of the central flux, we divide the sub-volumes
    into three groups: 
\begin{itemize}
  \item Quasar (Q) : 
    The central quasar dominates the entire sub-volume.
  \item Sub-dominant Quasar (QS) : 
    The central quasar dominates the central flux, but is
    not a significant fraction of the total flux of sources
    in the sub-volume. 
  \item Stellar (S): 
    The quasar flux in the central sources is smaller than
    the sum of stellar flux, and the entire central flux is
    small compared to the ionizing flux of the entire box.
    We note there can be two possible distributions of
    sources within the box: one being the presence of
    another major source that is not near the center of the
    sub-volume; the other being ionizing sources that are
    rather more uniformly distributed throughout the
    sub-volume. In our sub-volumes, the latter is more
    often to the case.
\end{itemize}

\section{Analytic Growth of Stromgren Spheres}

The growth of Stromgren spheres in a clumpy cosmological
    environment can be analytically modeled using
    \citep{2000ApJ...542L..75C},
\begin{equation}
  \frac{dR_{t}^{3}}{dt}=
     3 H(z) R_{t}^{3} + 
     \frac{3\dot{N}_{\rm ph}}
          {4\pi\left\langle n_{\rm H}\right\rangle}
     - C_{\rm H}
     \left\langle n_{\rm H}\right\rangle 
     \alpha_{\rm B}R_{t}^{3},
\end{equation}
    where the first term represents the Hubble expansion,
    and can be neglected in this context. The
    solution is straightforward
    \citep[eg,][]{1987ApJ...321L.107S}
\begin{eqnarray}
  R_{t} & = & R_{s}\left(1-\exp\left(-\frac{t}{t_{s}}\right)\right)^{-\nicefrac{1}{3}}\approx R_{s}\left(\frac{t}{t_{s}}\right)^{\nicefrac{1}{3}}\nonumber \\
  t_{s} & = & \left(C_{\rm H}\left\langle n_{\rm H}\right\rangle \alpha_{\rm B}\right)^{-1}
  \label{eq:free-streaming}\\
  R_{s} & = & \left(\frac{3\dot{N}_{\rm ph}}{4\pi C_{\rm H}\left\langle n_{\rm H}\right\rangle ^{2}\alpha_{\rm B}}\right)^{\nicefrac{1}{3}}.\nonumber 
\end{eqnarray}

We will use this analytic solution to compare to our numerical
    results. The IGM clumping factor $C_{\rm H}$ is
    calculated using a method appropriate for an SPH
    simulation, described by
    \citet{2009MNRAS.394.1812P},
\begin{equation}
  C_{\rm H}=\frac{\sum\rho_{i}^{2}h_{i}^{3}}{\left\langle \rho\right\rangle ^{2}\sum h_{i}^{3}},
\end{equation}
    where $h_{i}$ is the SPH smoothing length, and
    $\rho_{i}$ is the IGM density. The clumping factor in
    the entire sub-volumes has a value of $C_{H} \sim 4$,
    giving a recombination time about
    $t_{S} \sim \unit[100]{Myear}\gg t_{Q}$.  However we note
    the clumping factor can go up to $\sim 20$ within
    $\unit[1]{Mpc/h}$ radius from the center of the
    sub-volume, reducing $t_{S}$ to $\sim \unit[20]{Myear}$.

For the source flux, we make the simplest approximation,
    summing the flux of all sources with in the sub-volume
    to obtain one effective $\dot{N}_{\rm ph}$. We also
    assume that the volume consists of pure hydrogen (i.e.
    no helium, $X_{\rm H}=1$) in the calculation of the
    analytic model. The Case B recombination rate
    $\alpha_{\rm B}$ is taken from
    \citet{1997MNRAS.292...27H}, in consistency with the
    recombination rate table used in the radiative transfer
    simulation.

\section{Radiative Transfer Simulation}

\subsection{Ray-tracing Scheme}
For this study we have rewritten the {\small SPHRAY} Monte Carlo
    radiative transfer code \citep{2008MNRAS.386.1931A,
    2008MNRAS.388.1501C} in C with OpenMP(TM), so that it
    runs in parallel on shared memory systems ({\small
    P-SPHRAY}). We have also
    eliminated the on-the-spot approximation of
    recombination, so that instead recombination rays are
    emitted when the recombination photon deposit at a spot
    exceeds a threshold, as done in the code
    {\small CRASH} \citep{2003MNRAS.345..379M}. The parallel version
    allows us to trace many rays at the same time on several
    CPUs within a single time step, achieving a near-linear
    speed up with a small number (10 $\sim$ 30) of
    CPUs. Monochromatic rays are sampled from frequency
    space according to the source spectrum.  Hydrogen and
    Helium ionization and recombination are both traced.
    The tabulated atomic reaction rates of
    \citet{1997MNRAS.292...27H} taken from the serial
    version {\small SPHRAY} are used. 

As a new addition to the code we model the secondary
    ionization of high energy photons using a fit to the
    Monte Carlo simulation results of
    \citet{2010MNRAS.404.1869F, 1985ApJ...298..268S}. These
    fast non-thermal-equilibrium electrons produced from
    high energy ionization photons (e.g., X-ray photons) may
    collide and secondarily ionize more neutral atoms,
    increasing the ionizing efficiency of the quasar sources
    by allowing one high energy photon to ionize tens of
    neutral atoms. On the other hand, our current
    post-processing approach is incapable of modeling the
    heating from the residual energy of the ionizing
    photons, for the thermal evolution (hydro) is decoupled
    from radiative transfer.  We note that the local heating
    of ionizing photons can be thought of as part of the
    local thermal coupling between the liberated luminosity
    and the surrounding environment described in Section 2.
    The long range heating of the harder photons is not
    directly modeled. 

\subsection{Parametrization of the Ionized Bubble}
\label{sub:Parameterizing-the-Ionization}
For this study, we are more interested in the ionized
    bubbles located at the centers of sub-volumes, because
    they are associated with the main halos in the sub-volumes.
    We will call these central ionized bubbles the central
    ionized regions (CIR).

When quantifying the size of a CIR we measure the
    spherically averaged fraction of a species, $x$ using
    (we show the example of HII regions, similar definitions
    hold for other species):
\begin{equation}
  x_{\rm HI}(r) = 
     \frac{\int\delta(r'-r)x_{\rm HI}\rho dV}
          {\int\delta(r'-r)\rho dV}
          \doteq \frac{\sum_{\rm shell}x_{\mathrm{HI}, i}m_{i}}
              {\sum_{\rm shell}m_{i}}.
\end{equation}
    Here the sums are carried out by binning SPH particles
    (index $i$) in shells. Particles are assigned as a whole
    to shells by their center position rather than being
    integrated over the SPH kernel with in the shell. We
    define the averaged CIR radius at a given threshold
    $x^{*}$ to be the first crossing of $x(r)$ with $x^{*}$,
    or 
\begin{equation}
  \hat{R}_{S}(x)=\inf\{r|x_{\rm HI}(r)=x^{*}\}.
\end{equation}
    The first crossing corresponds to the edge of the CIR,
    and later crossings indicate the edges of the nearby
    ionized regions. The motivation for this definition
    comes from its similarity to that used to quantify
    ionized regions seen in quasar absorption line spectra,
    the Ly$\alpha$ absorption Near-Zone radius $R_{NZ}$ by
    \citet{2006AJ....132..117F}.  In practice, we take as
    $\hat{R}_{S}$ the average of the two nearest bins
    $r_{L}$ and $r_{R}$ which satisfy
    $[x_{\rm HI}(r_{L})-x][x_{\rm HI}(r_{R})-x]<0$. 

The spherically averaged radius $\hat{R}_{S}$ is a simple
    quantity to use in comparisons but the angular
    variations in the properties of the CIR are completely
    lost. To quantify the angular dependency, we bin the SPH
    particles into angular cones and calculate the averaged
    fraction within the cones, 
\begin{equation}
  x_{\rm HI}(r,\theta,\phi) \doteq 
  \frac{\sum_{\rm cone,shell} x_{\mathrm{HI}, i} m_{i}}
  {\sum_{\rm cone,shell}
  m_{i}}.
\end{equation}
    The angular dependent CIR radius is
\begin{equation}
  R_{s}(x,\theta,\phi)=\inf\{r|x_{\rm HI}(r,\theta,\phi)=x^{*}\}.
\end{equation}
    Note that in general
    $<R_{s}(x,\theta,\phi)>\neq\hat{R}_{S}$, because
    averaging and binning do not commute.  For measuring
    $R_{s}(x,\theta,\phi)$, we use $12\times16^{2}$ cones.
    We note that increasing to $12\times32^{2}$ cones does
    not significantly alter the results.

We measure the anisotropy of the ionized bubble from the
   variance of $R_{s}$
\begin{equation}
  \label{eq:anisotropy}
  A_{s}(x)=ST{}_{k}[R_{s}(x,\theta_{k},\phi_{k})],
\end{equation}
    where $ST_{k}$ stands for the standard derivation. We
    run the simulation with sufficient number of rays so
    that the typical shot-noise contribution towards $A_{s}$
    is negligible, as shown in the next section.

Three different levels of species fraction $x$ are used to
    define three different levels of CIR fronts in this study:
\begin{itemize}
  \item  
    Inner front, $x=0.1$ for the neutral fraction, or
    $x=0.9$ for the ionized fraction; 
  \item 
    Middle front, $x=0.5$ for the neutral fraction; 
  \item 
    Outer front, $x=0.9$ for the neutral fraction, or
    $x=0.1$ for the ionized fraction. 
\end{itemize}

The Inner front corresponds to the near-ionized edge of the
    Stromgren sphere, and the Outer front corresponds to the
    near-neutral edge of the Stromgren sphere.  We however
    note that these choices are for illustrative purposes
    and they do not have any direct correspondence with
    threshold values used to detect $\unit[21]{cm}$ or
    Ly$\alpha$ observation signatures.

\subsection{Shot-noise and Convergence}

One issue which must be addressed in Monte Carlo ray
    tracing (RT) schemes is the presence of shot-noise and
    its effect on convergence of results, something
    particularly important when sampling is also in
    frequency space \citep[eg, ][comments on
    {\small CRASH}]{2006MNRAS.371.1057I}. Shot-noise artificially
    increases the angular anisotropy $A_{s}$ measure, and so
    is of direct concern for this study. 

We define a shot-noise parameter $\gamma$ to be the ratio
    between the number of photons in a ray packet,
    $n_{r}^{0}$ and the number of atoms in an SPH particle
    $n_{p}^{0}$, 
\begin{equation}
  \label{eq:Shot-Noise-Parameter}
  \gamma=\frac{n_{r}^{0}}{n_{p}^{0}}.
\end{equation}
    A small $\gamma$ guarantees that the ionization front
    cannot advance by more than one particle in one time
    step. When $\gamma\ll1$, the ionization front advances
    slowly and the shot-noise is controlled.  One of course
    still needs to ensure the rays have a sufficient angular
    resolution to resolve the angular scale used in the
    calculation of the anisotropy. 

In MassiveBlack, a typical SPH gas particle has a mass of
    $\unit[5\times10^{7}]{M_{\odot}/h}$, equivalent to
    $n_{p}^{0}=9\times10^{64}$. For ray tracing with
    $10^{5}$ steps and $128$ rays per time step,
    ($1.28\times10^{7}$ rays), and with the total luminosity
    listed in Table \ref{tab:Total-ionization-photons}, an average
    packet contains $n_{r}^{0}=3\times10^{63}$ photons. The
    shot-noise parameter is therefore $\gamma=0.03\ll1$ for
    our typical runs.

We can also empirically confirm that convergence is reached
    by increasing the number of rays used in the simulation
    and showing that the quantity of interest is insensitive
    to further increases. We perform such a convergence test
    with 3 runs on sub-volume 4 with (i) $1.3\times10^{8}$
    rays, $\gamma=0.003$, (ii) $1.3\times10^{7}$ rays,
    $\gamma=0.03$, and (iii) $1.3\times10^{6}$ rays,
    $\gamma=0.3$. We quantify the similarity of the
    ionization front using the correlation coefficient of
    $R_{s}(\theta,\phi)$ between runs, shown in
    Table \ref{tab:Correlation-Coefficients.}.  The higher
    correlation between the runs with more rays indicates
    that the simulation has effectively converged.
    Therefore, for the runs we use $1.3\times10^7$ rays.

\begin{table}
  \begin{tabular}{cccc}
  \toprule 
  Runs & HII Inner & HII Middle & HII Outer\tabularnewline
  \midrule
  \midrule 
  $\gamma = 0.003 ~\mbox{and}~ 0.03$ & 0.96 & 0.98 & 0.86\tabularnewline
  \midrule 
  $\gamma = 0.03 ~\mbox{and}~ 0.3$  & 0.74 & 0.94 & 0.68\tabularnewline
  \bottomrule
  \end{tabular}
  
  \caption{Correlation coefficients in convergence test for
    number of rays. See Equation
    \ref{eq:Shot-Noise-Parameter} for definition of $\gamma$.
    Shown in the table are the correlation coefficient of
    the center bubble radius $R_{s}(\theta,\phi)$ between
    different runs.}
  \label{tab:Correlation-Coefficients.}
\end{table}

\section{Results and Discussion}

\subsection{Uniform Density Field}
\begin{figure}
  \includegraphics[width=1\columnwidth]{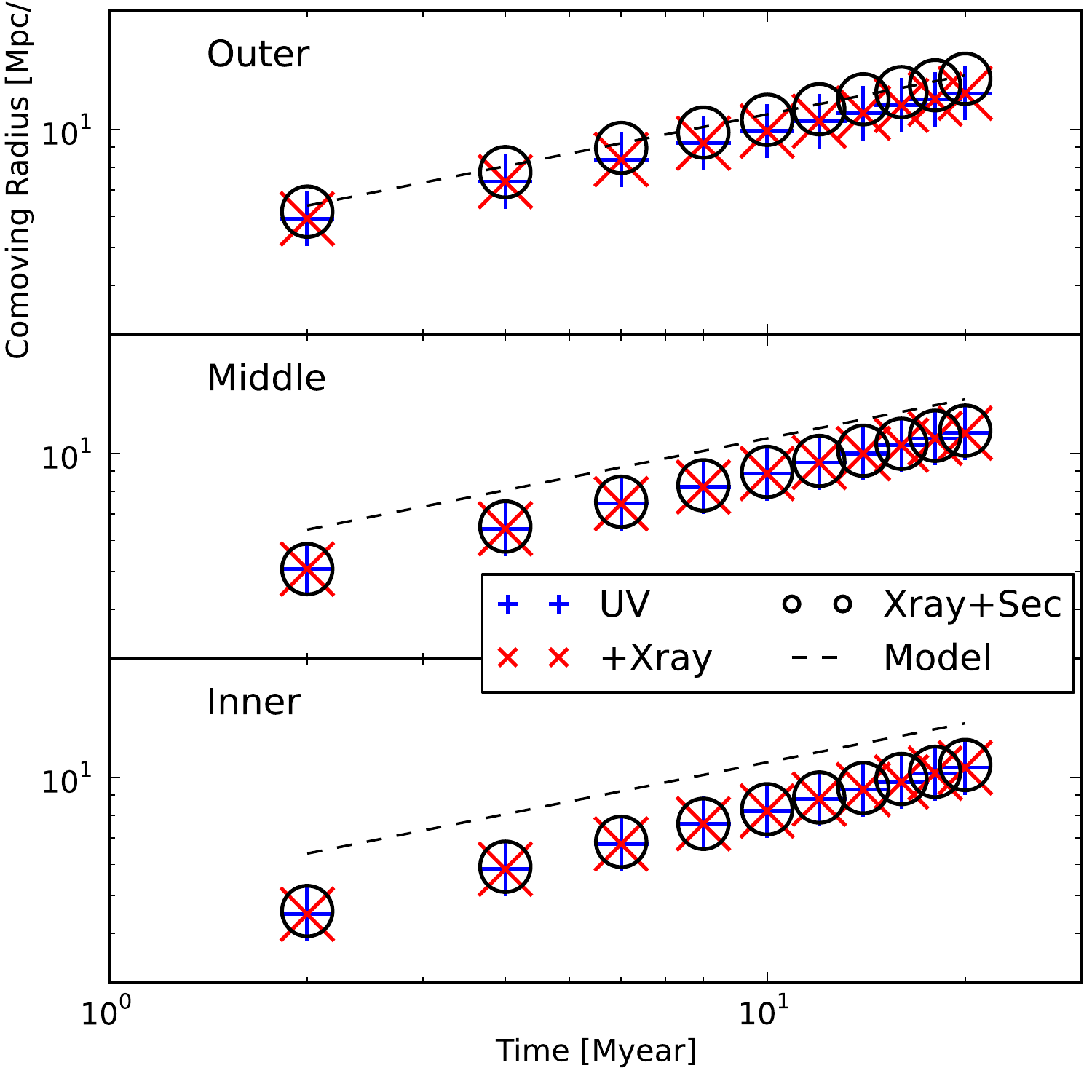}

  \caption{CIR Growth in a Uniform Density Field. 
    The symbols are:
    (i) cross[+]: UV only; (ii) cross[x]: UV and X-ray (iii)
    circle: UV and X-ray, with Secondary ionization.  The
    analytic model (dashed lines) neglects Helium and
    assumes $X_{H}=1.0$.}
  \label{fig:Growth-in-a}
\end{figure}

We first test {\small P-SPHRAY} with a source in a uniform density
    field at $z=8$, with H number density
    $n_{\rm H}=\unit[2\times10^{-4}]{cm^{-3}}$, and
    uniform temperature $T=\unit[10^{4}]{K}$. The hydrogen
    mass fraction is $X_{\rm H}=0.76$, the cubical box
    has a side length of $\unit[50]{Mpc/h}$, and we evolve
    the radiation and ionization fractions for a duration of
    $\unit[2\times10^{7}]{yrs}$. We carry out 3 separate
    simulations, considering different central source
    properties for each one as follows: (i)
    $\unit[1.2\times10^{56}]{sec^{-1}}$ UV; (ii)
    $\unit[1.2\times10^{56}]{sec^{-1}}$ UV,
    $\unit[0.3\times10^{55}]{sec^{-1}}$ soft X-Ray,
    $\unit[0.5\times10^{55}]{sec^{-1}}$ hard X-Ray; (iii)
    same as (ii), with secondary ionization.  The
    luminosity of the source is motivated by the luminosity
    of the center sources in sub-volume 0.  The growth of the
    CIR fronts in the simulations are shown as a function of
    time in Figure \ref{fig:Growth-in-a}.
    We show as separate panels the behaviors of the three
    parts of the ionization front: inner, middle and outer
    (as defined in Section \ref{sub:Parameterizing-the-Ionization}). 

There are three highlights from the uniform density field
    simulations: (i) The fronts in general agree well with the
    analytic model described in \ref{eq:free-streaming}.
    (ii) There is a smooth transition from neutral to
    ionized state at the front, indicated by the
    $~\unit[1]{Mpc/h}$ difference between the inner front
    and the outer front. The transition is mostly due to the
    penetration of harder UV photons. (iii) The effect of
    X-ray photons and secondary ionization is more evident
    on growth of the outer front than of the inner and
    middle fronts.  This is because the secondary
    ionization affects the way the harder photons interact
    with the matter the most, and harder photons travel
    further into the neutral region.  We note that by adding
    in the effect of secondary ionization increases the
    outer radius of ionized bubbles by approximately 10\%.

\subsection{MassiveBlack: Visualization}

\begin{figure*}
  \includegraphics[width=0.49\textwidth]{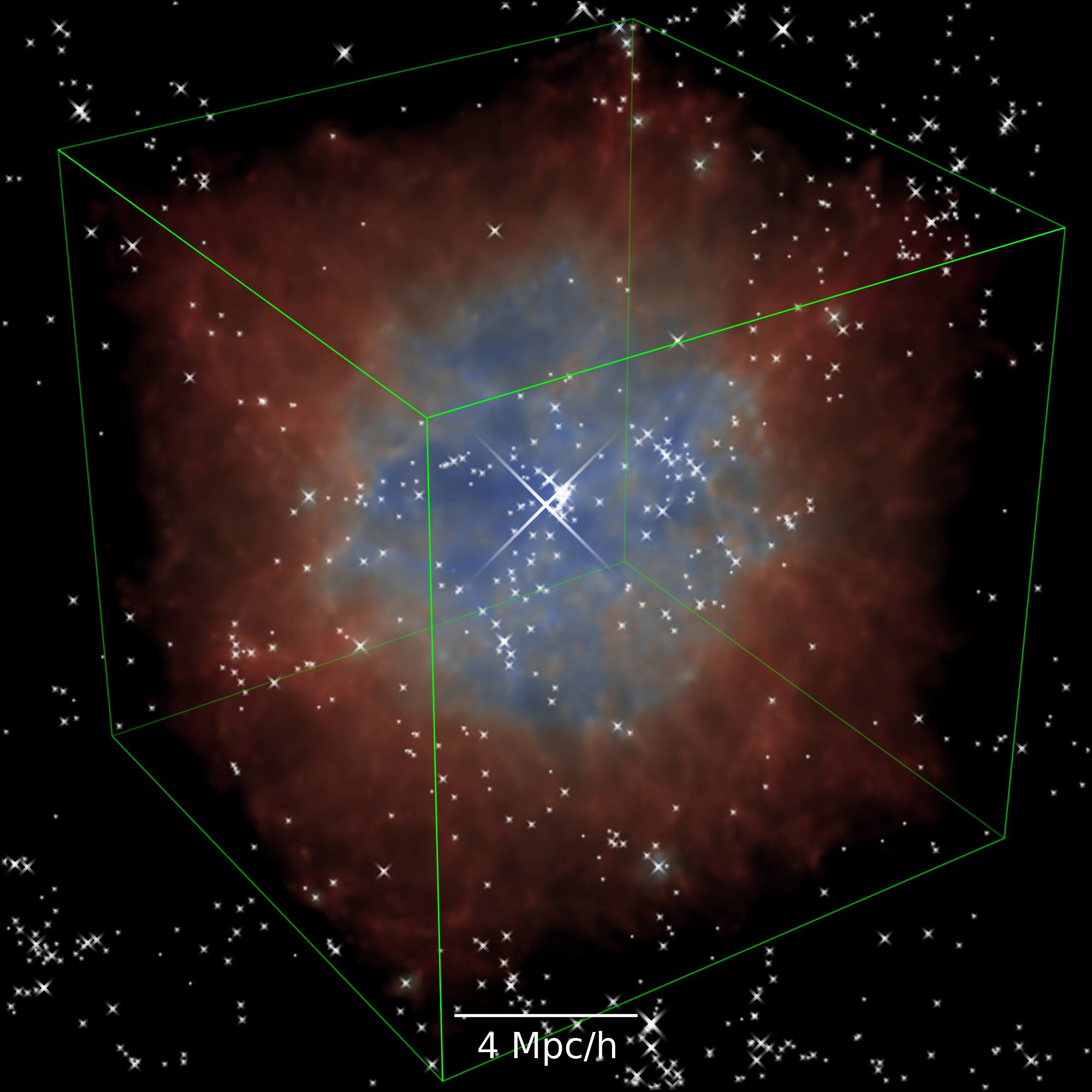}
  \includegraphics[width=0.49\textwidth]{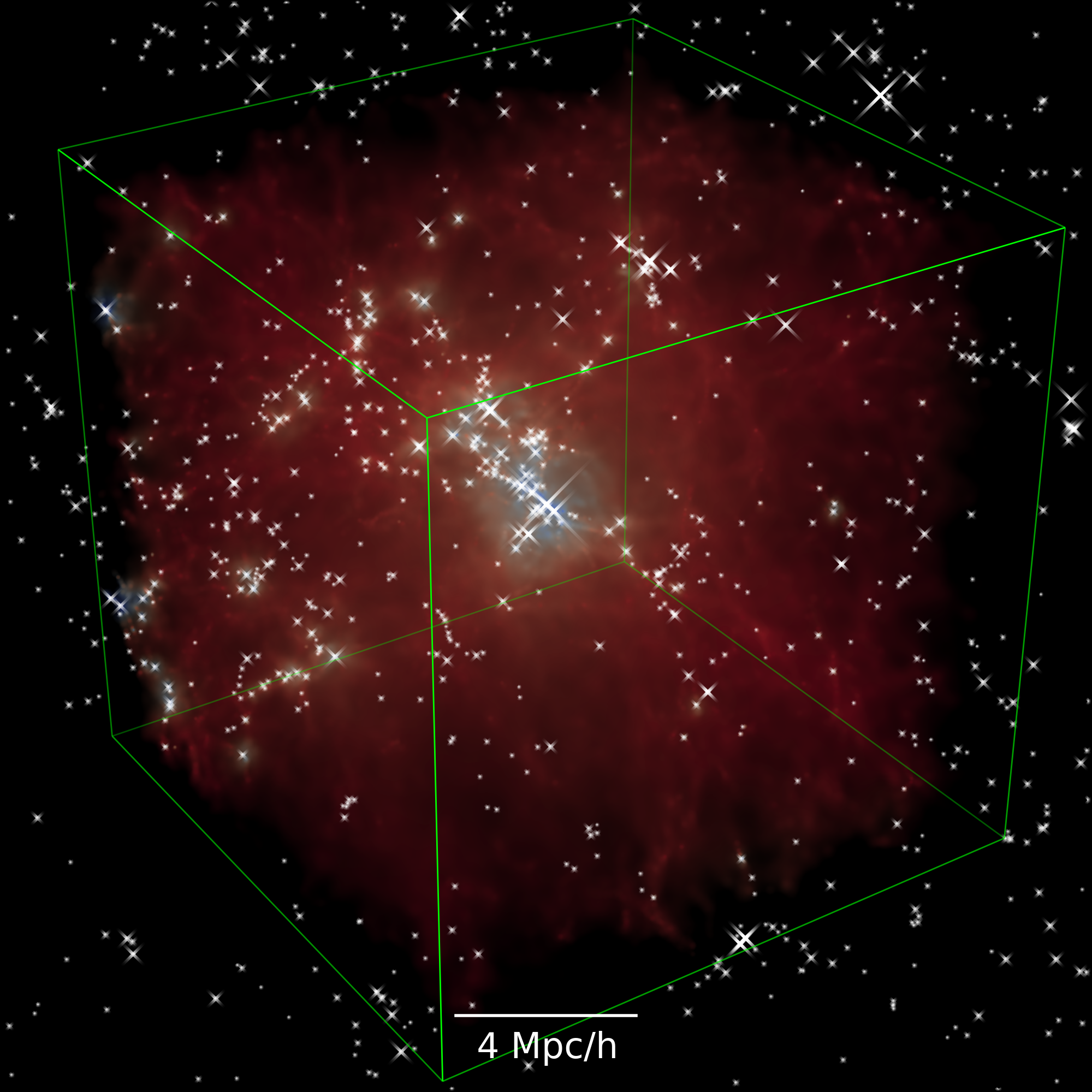}

  \caption{Q (quasar, left) type and S (stellar, right) type
  ionized bubbles in MassiveBlack. Both panels are of
  comoving length $\unit[15]{Mpc/h}$ per side.  A camera is
  put at about $\unit[60]{Mpc/h}$ from the center of the
  sub-volume and a perspective transformation is applied to
  form a projection onto the image plane of the camera.  Red
  color corresponds to fully neutral IGM and blue color
  corresponds to fully ionized IGM. Yellow is in between the
  two states.  Crosses are the sources; both stellar and
  quasar sources in the entire sub-volume are shown. It is
  interesting to notice the lack of a major source
  (compared to the bright quasar in sub-volume 3) in
  sub-volume 4.}
  \label{fig:Q-vs-S}
\end{figure*}

We show the CIR bubble of the most extreme Q type and S type
    sub-volumes (sub-volume number 3 and 4 in
    Table~\ref{tab:Total-ionization-photons} respectively)
    in Figure \ref{fig:Q-vs-S}.  The halo mass of both are similar,
    $\unit[60\times10^{10}]{M_{\odot}h^{-1}}$ and so are
    their total star formation rates. However sub-volume 3
    has a large ionized HII region associated with the
    bright active quasar in the center of the sub-volume,
    whilst in sub-volume 4 the HII region formed merely from
    stellar sources are small. Note that, even though in a 
    comparable mass halo, the central black hole is much
    smaller in sub-volume 3 and hence its activity has a
    negligible impact. This example aims to illustrate how,
    in the presence of an active early quasar, the ionized
    bubble is far more extensive than one driven by star
    formation alone.

\begin{figure*}
    \hfill{}%
    \includegraphics[width=0.2\textwidth]{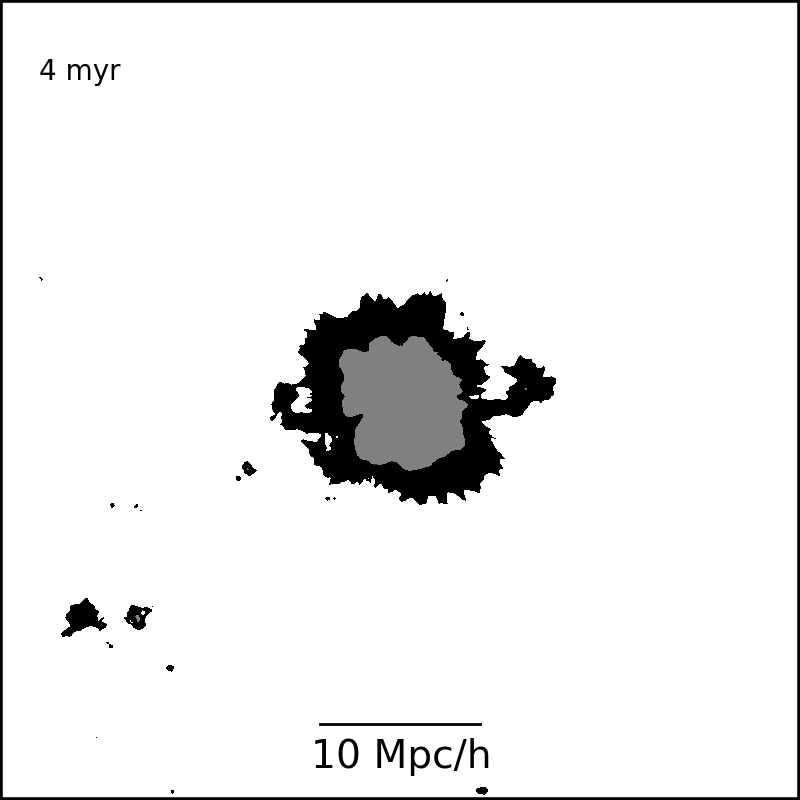}%
    \includegraphics[width=0.2\textwidth]{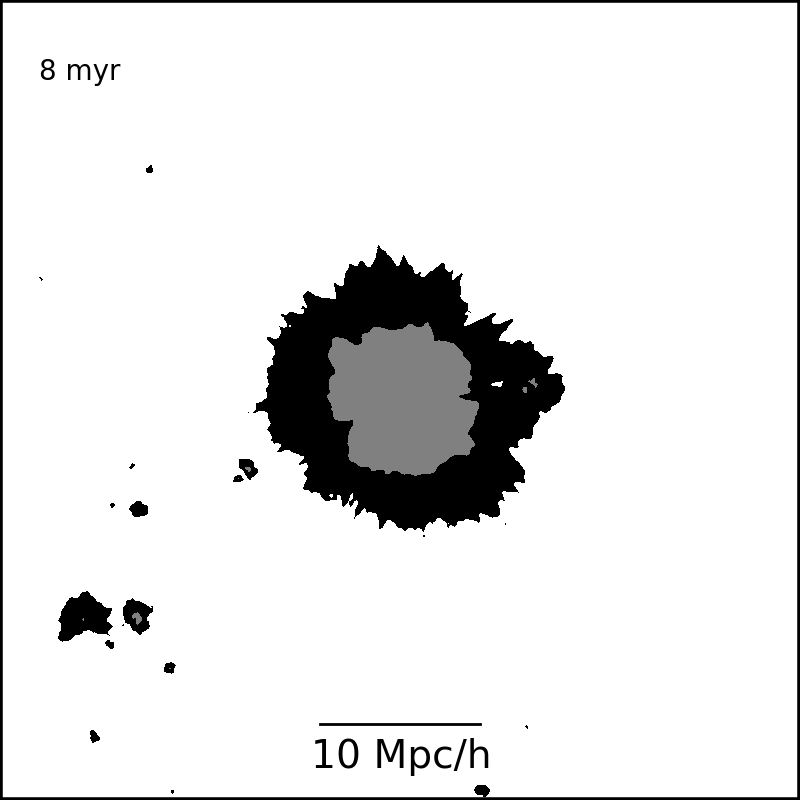}%
    \includegraphics[width=0.2\textwidth]{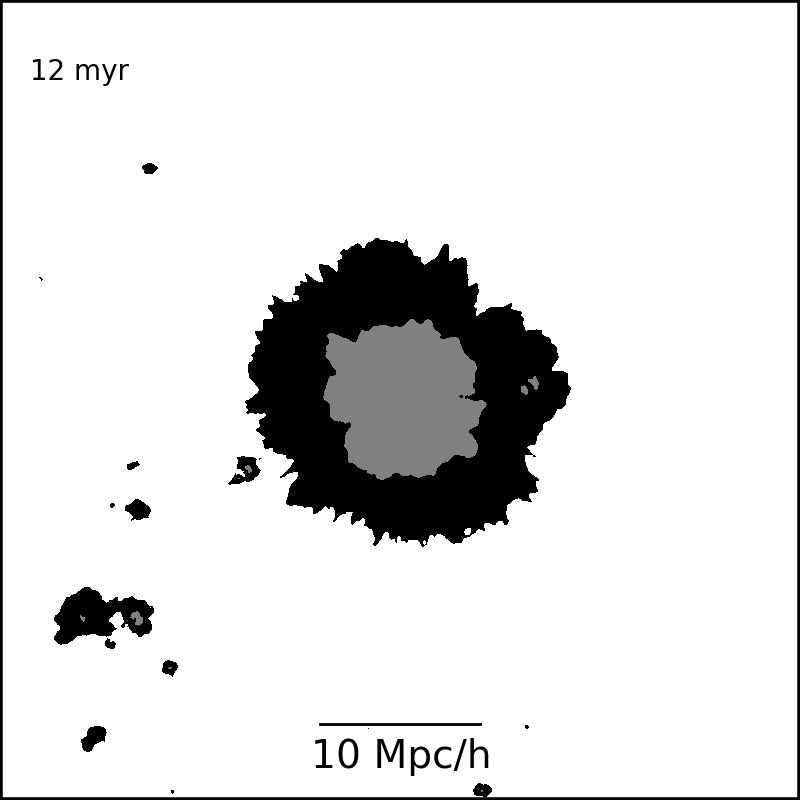}%
    \includegraphics[width=0.2\textwidth]{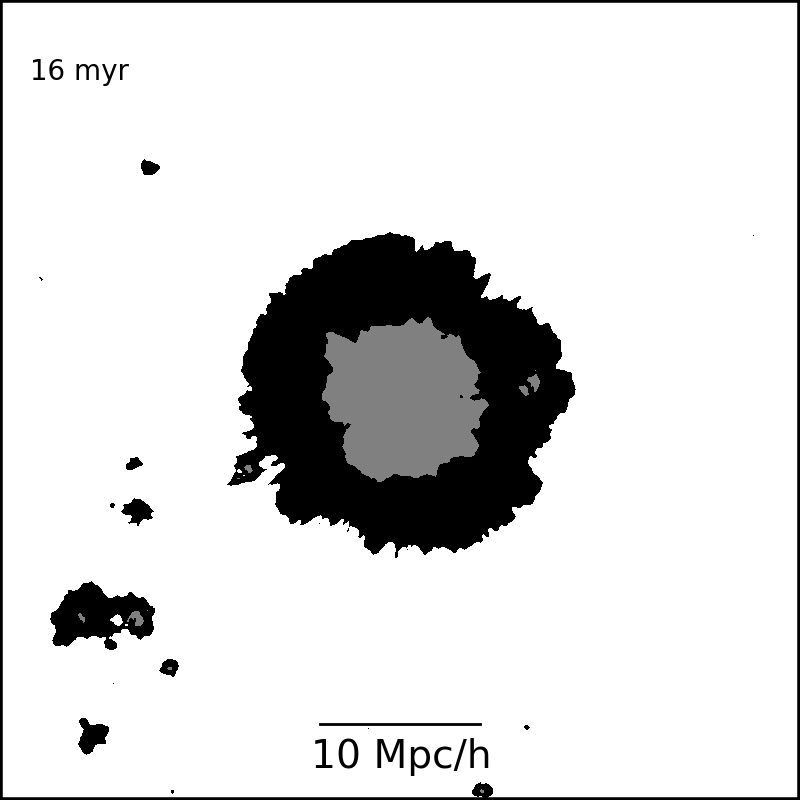}%
    \includegraphics[width=0.2\textwidth]{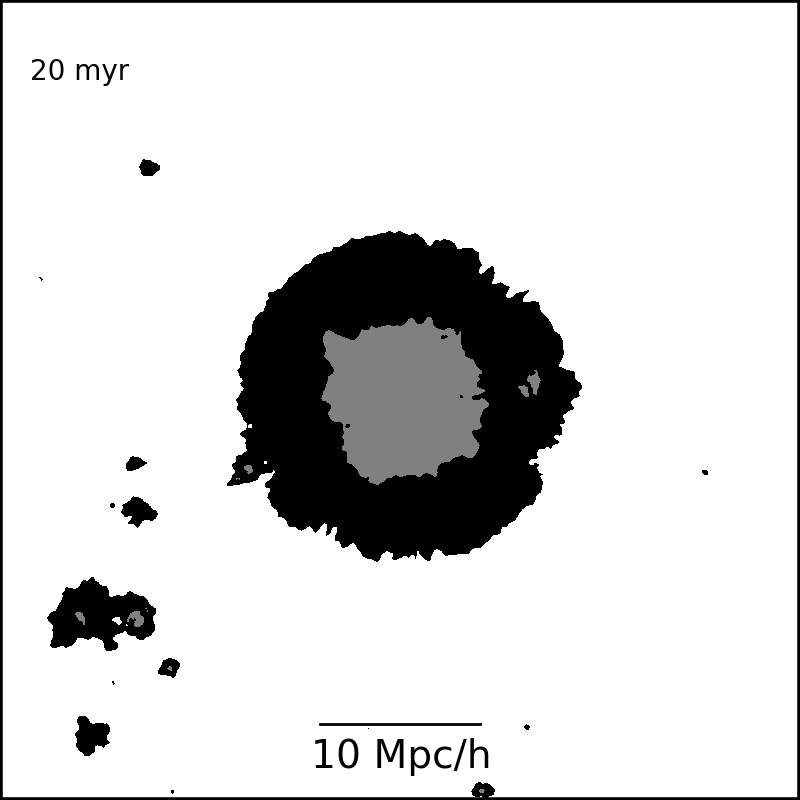}%
    \hfill{}

    \caption{Evolution of the ionization front.  
    The evolution of the ionization front in sub-volume \#0
    is shown. The ionization
    front ($x_{\rm HI} \in [0.1,\:0.9]$) is marked in
    black, and the fully ionized near
    zone ($x_{\rm HI}<0.1$) is marked with gray.}
    \label{fig:Evolution-of-if}
\end{figure*}

The time evolution of the HII bubble in sub-volume 0, where
  three halos coexist, is shown graphically with a slice
  through the center of the simulation volume every
  $\unit[2\times10^{6}]{yr}$ in Figure
  \ref{fig:Evolution-of-if}.  It is interesting to observe
  that a neighboring bubble on the right merges with the
  center bubble and this increases the size of the outer
  front in that direction, contributing to the anisotropy of
  the front.  We investigate this issue in the next section
  in a more quantitative way.

\subsection{MassiveBlack: Spherically Averaged Radius of Ionized Regions}

\begin{figure}
  \includegraphics[width=1\columnwidth]{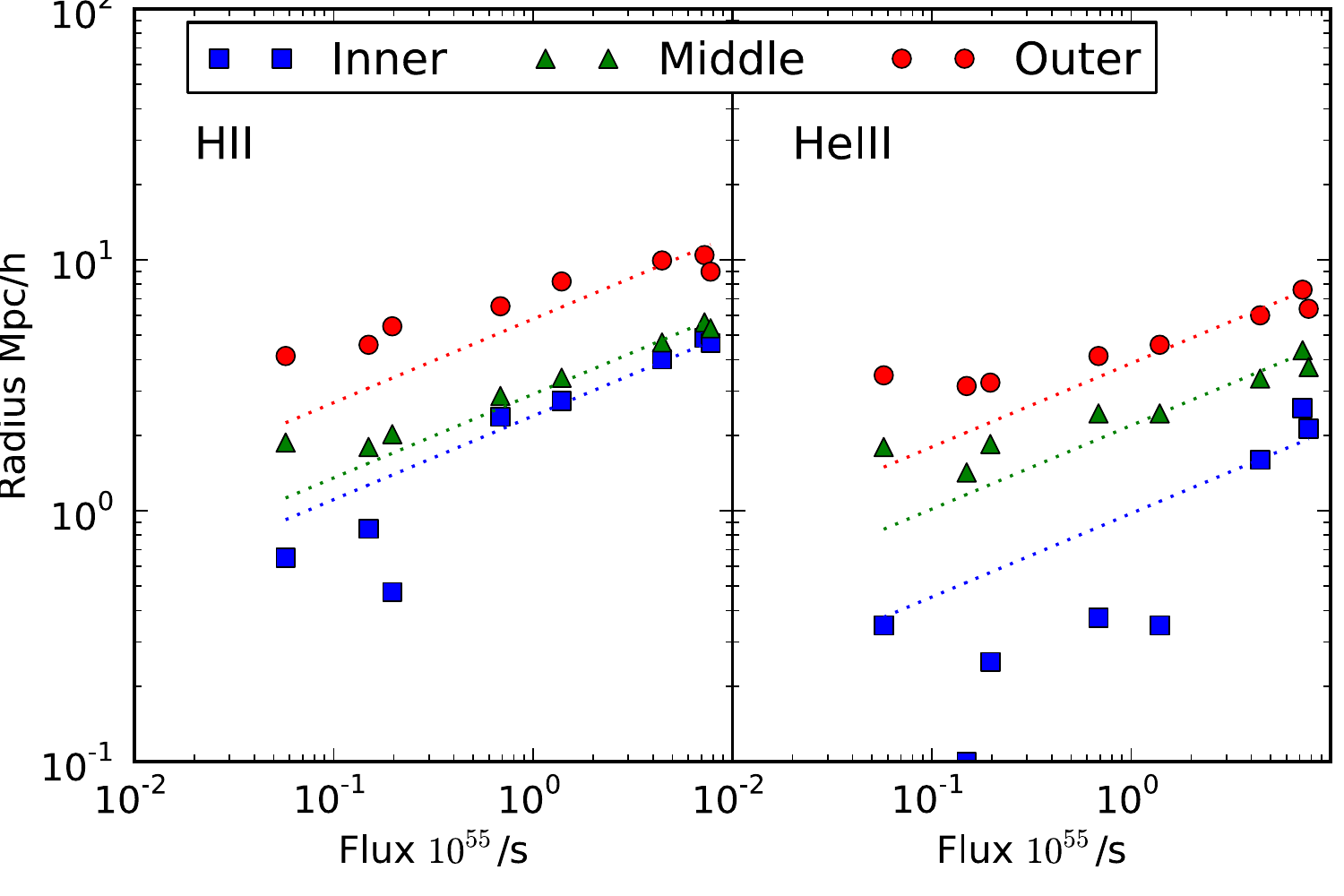}

  \caption{HII and HeIII CIR Front Radius. 
  The left panel is for the HII CIRs, the right panel for
  the HeIII CIRs. The horizontal axis is the central source
  flux. The dotted line in the first two panels shows the
  scaling assuming the analytic model in Equation
  \ref{eq:free-streaming}.}
  \label{fig:HII-and-HeIII}
\end{figure}

We now examine the results of the radiative transfer
    post-processing of the eight sub-volumes from
    MassiveBlack.  The averaged CIR radius $\bar{R_{s}}$ for
    HII and HeIII, as defined in Section
    \ref{sub:Parameterizing-the-Ionization}, are shown in
    Figure \ref{fig:HII-and-HeIII} as functions of the
    central source flux. We fit the growth against the
    scaling relation described by Equation
    \ref{eq:free-streaming}, and find that at the low
    luminosity end (mainly S type and QS type sub-volumes)
    there is a significant deviation from the fit for the
    inner and middle fronts. 
    The outer front is more extended than the simple uniform
    density field simulation, albeit given the similar
    source spectra, hinting that the structure in the IGM
    is also contributing to the smoothing of the front. This
    smoothing is a phenomenon similar to that described by
    \citet{2007MNRAS.374..960W}.  We also attribute the
    effect to the clustering of sources; however unlike the
    smoothing due to secondary ionization, the clustering
    contribution is anisotropic, and we discuss it in
    Section \ref{sub:Anisotropy}.

\begin{figure}
  \centering\includegraphics[width=0.8\columnwidth]{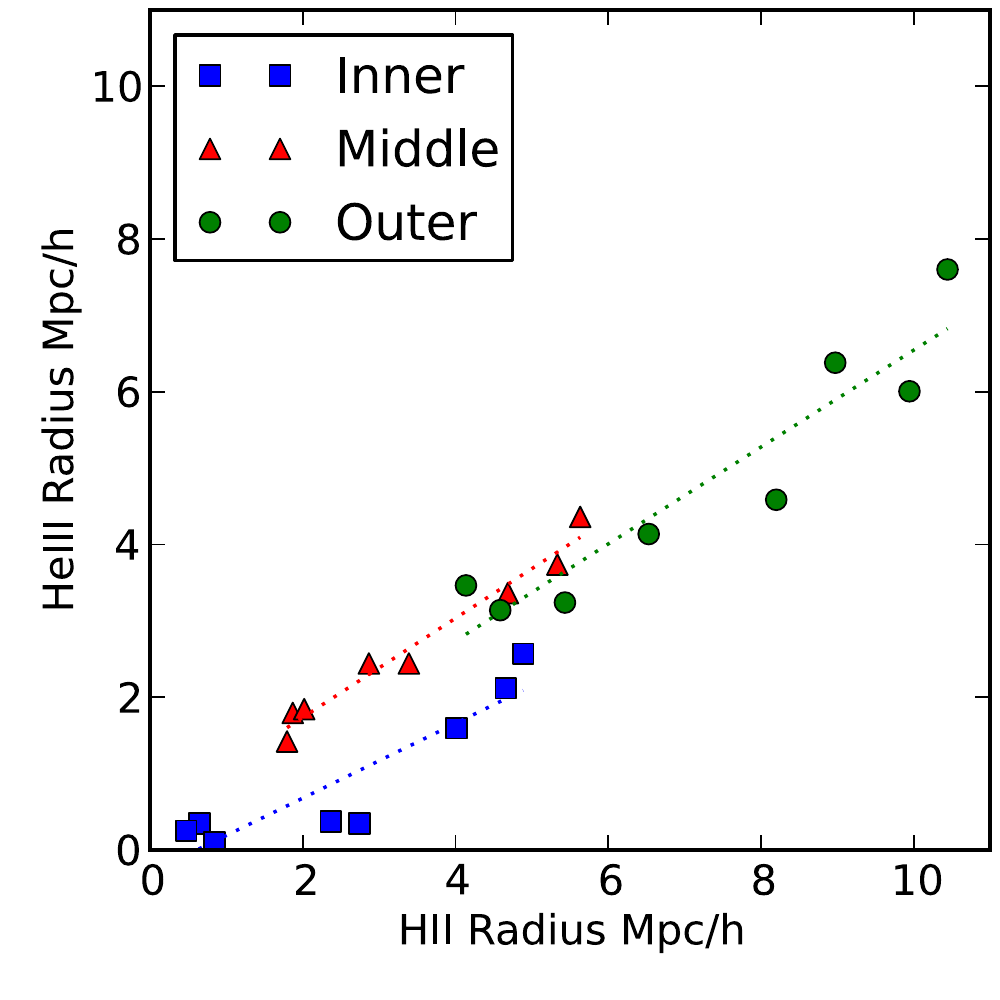}

  \caption{ Correlation between HeIII CIR and HII CIR.
  HeIII CIRs is plotted against HII CIRs, and the linear
  regression for three fronts are shown with the dotted
  line.}
  \label{fig:Correlation-HeIII-HII} 
\end{figure}

In Figure \ref{fig:HII-and-HeIII}, we can see that the HII and HeIII
    CIR radii appear to grow together. The correlation
    between the two can be seen directly by plotting one
    against the other, which we do in
    Figure \ref{fig:Correlation-HeIII-HII}, where we find there is a
    strong correlation between the HII radius and HeIII
    radius. The HeIII radius is smaller than the HII radius,
    agreeing with the finding of \citet{2012MNRAS.421.2232F}
    who used the ray tracing code {\small C2RAY}. We note however
    that the treatment of secondary ionization and Helium
    ionization in {\small P-SPHRAY} is similar to that of
    {\small C2RAY}, and
    an agreement is not surprising.

\begin{figure*}
  \includegraphics[width=1\textwidth,trim=40pt 0 0 0,clip]{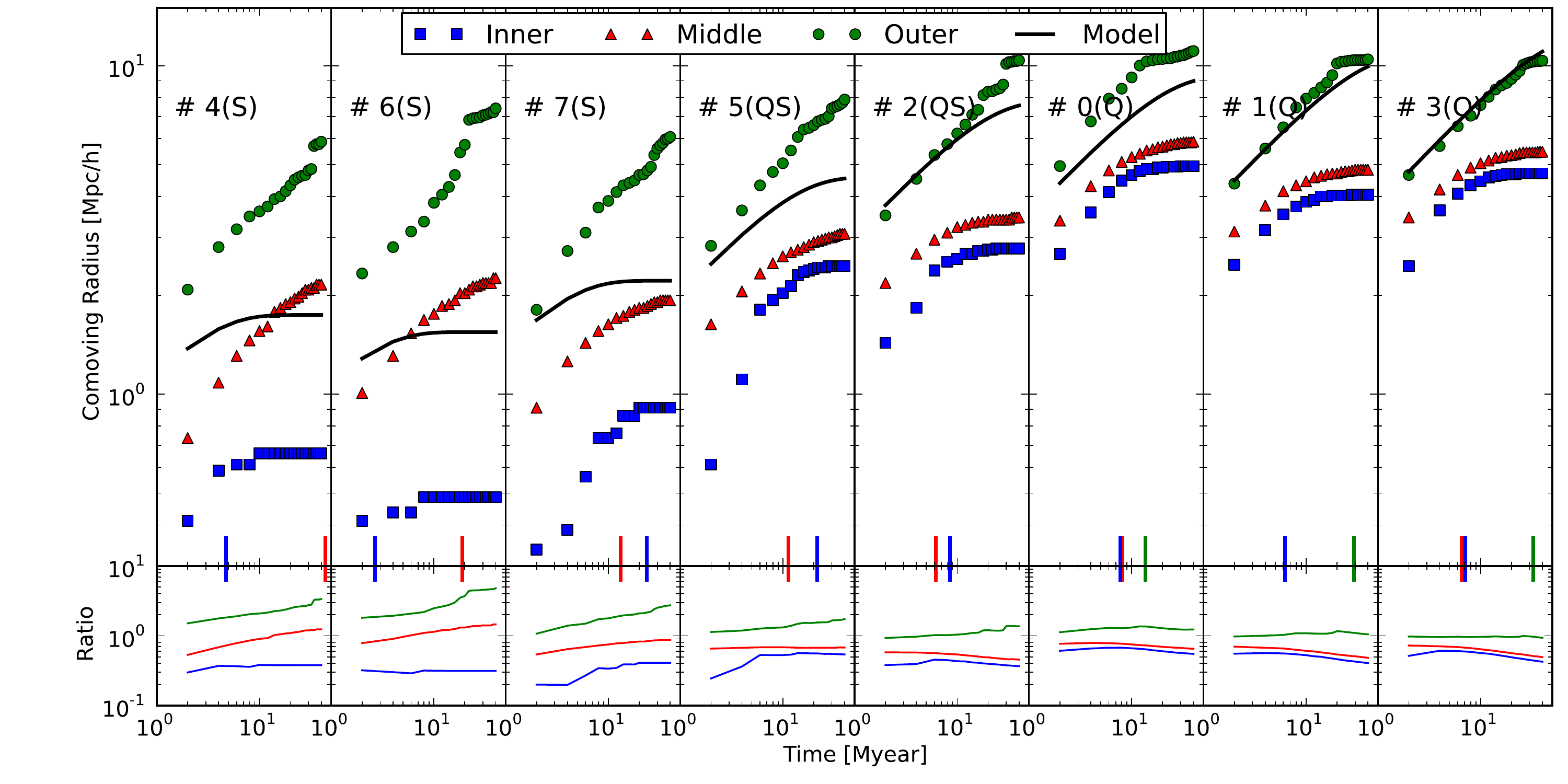}

  \caption{ Evolution of three quasar driven sub-volumes.
  The top panels show the CIR radius as function of time;
  the thick black line is the fiducial analytic model as
  described in the text (Equation \ref{eq:free-streaming}).
  The vertical short lines show the recombination time in an
  analytic model as if it is fitted as a parameter to the
  simulation data points, which are marked with squares
  (Inner), triangles (Middle), and circles (Outer). The
  bottom panels show the ratio between the radius and the
  fiducial model.}
  \label{fig:Evolution-of-three}
\end{figure*}

The time evolution history of the HII CIR radius for all
    sub-volumes is shown in Figure \ref{fig:Evolution-of-three}. We
    compare the evolution of the three parts of the fronts
    with the prediction of the analytic model (in equation
    \ref{eq:free-streaming}), in which we have used the
    clumping factor and mean H density measured within the
    final CIR in the simulation to estimate a fiducial
    recombination time $t_{S}$.

The growth of the CIR in the Q type sub-volumes flattens off
    much earlier than one might expect from a free streaming
    law. In order to ascertain the relevant physical
    timescales, we fit the time evolution of the fronts to
    the full analytic model in \ref{eq:free-streaming}, and
    extract the effective recombination time $\hat{t}_{S}$
    as a fit parameter, then mark the time in the plot. We
    can see that for the quasar driven (type Q) sub-volumes,
    all three fronts (inner, middle and outer) growth stop at
    about $\unit[10]{Myear}$, on the same order of the life
    time of the quasar $t_{Q} \sim \unit[20]{Myear}$ and the
    fiducial recombination time $t_{S} \sim \unit[20]{Myear}$.
    The deviation from free streaming indicates that by
    merely increasing the life span $t_{Q}$ of the central
    quasar, we can not substantially increase the size of
    their CIR.

In S type type sub-volumes, there is a similar tailing off
    for the inner fronts. However, the outer fronts continue
    their free streaming growth, behaving differently from
    the analytic model, which has stopped due to reaching
    the recombination time.  We attribute this apparent
    excessive growth of the averaged S type outer front to
    the anisotropic growth via merging with other small
    ionized bubbles that are close to the CIR, as described
    later in the paper.

We note that our Monte-Carlo RT approach allows for
    superluminal growth of the CIR.  This nonphysical
    situation happens well before the first snapshot time
    which is 2 million years after the sources are turned
    on. The first snapshot, in which the Stromgren sphere
    typically has grown to about 10\% of the full size,
    gives an upper bound to the contribution from
    superluminal growth. We conclude that the superluminal
    growth occurring at early time ($< \unit[2]{Myear}$) and
    small radii ($<10\%$) does not play an important role in
    the final shape of the ionized bubble.  We refer the
    readers to \citet{2006ApJ...648..922S} for a more
    detailed discussion of the role of the nonphysical
    superluminal phase of the growth of an ionization front.

\subsection{MassiveBlack: Anisotropy of Ionized Regions}
\label{sub:Anisotropy}
\begin{figure}
  \includegraphics[width=1\columnwidth]{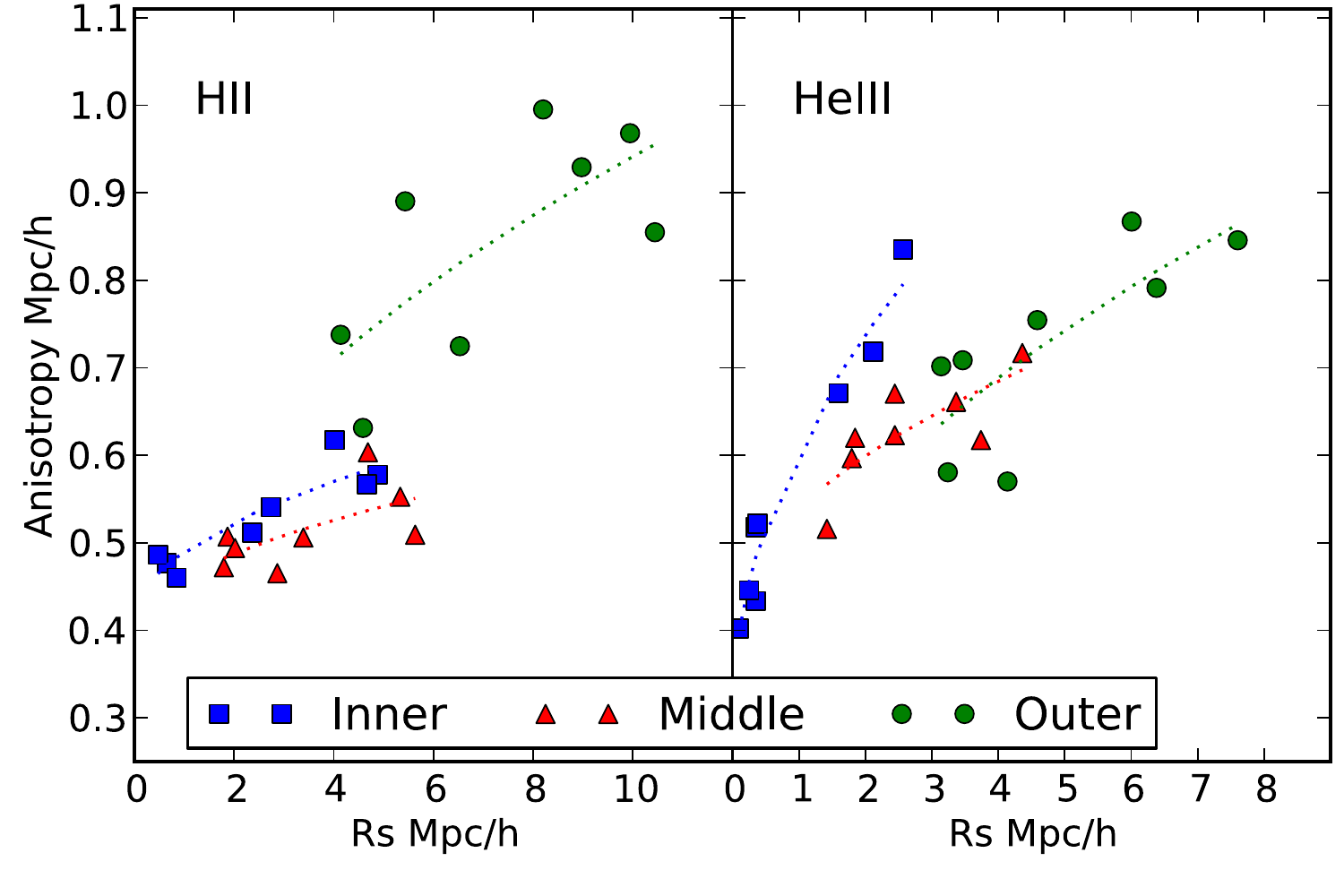}

  \caption{Anisotropy .vs. Radius. 
  The left panel shows the anisotropy of three (inner,
  middle, outer) HII CIR fronts defined as in
  Equation \ref{eq:anisotropy}. The right panel shows the anisotropy
  of the HeIII fronts. The dotted lines are fits against a
  square-root + linear offset model and are only meant to
  guide the eye. All 8 sub-volumes are displayed.}
  \label{fig:Anisotropy-and-Radius.}
\end{figure}

In Figure \ref{fig:Anisotropy-and-Radius.} we show the anisotropy
    measured using Equation \ref{eq:anisotropy} of all 8
    sub-volumes.  The anisotropy of different ionization
    fronts as a function of radius are marked with different
    symbols. As a reminder, in Equation \ref{eq:anisotropy},
    the anisotropy $A_{s}$ is the standard deviation of the
    ionization front radius, so that by comparing to the
    radius plotted on the x-axis it can be seen that the
    fronts are not very far from spherical symmetry
    ($ \sim $ 10\% variations).  We find that larger
    ionized regions (corresponding to brighter quasar
    sources) are in general associated with more anisotropy. 

In the simulation, the anisotropy of the inner and middle
    radii of the HII regions do not significantly depend on
    the radius (the curve is relatively flat), but the outer
    fronts have more anisotropy.  The HeIII regions show a
    similar feature, except for the inner fronts of three
    type Q sub-volumes, which have significantly stronger
    anisotropy.

These phenomena lead us to the following explanation,
    incorporating two contributing mechanisms:
\begin{enumerate}
  \item 
    The anisotropic distribution of gas that attenuates the
    ionizing photons; when the density is high in a
    particular direction, the extra absorption decreases the
    radius of the ionized region in that direction.
  \item 
    The merging of nearby bubbles from clustered halos; when
    the density is sufficient to host bright sources lying
    in a certain direction, their extra photo-ionization
    increases the bubble radius in that direction. The outer
    front is more sensitive to merging than the inner and
    middle front.
\end{enumerate}
    For a small CIR (or a CIR in its early growing stage), no
    merging has occurred and only the density induced
    anisotropy is present.  As the CIR grows, it overlaps
    with nearby ionized regions, resulting in the second type
    of anisotropy (due to overlapping).  We note that this
    does not contradict the finding that clustering
    contributes to the smoothness of the extended ionization
    front \citep{2007MNRAS.374..960W}, because by
    definition, after spherical averaging, the existence of
    surrounding bubbles will result a smoother averaged
    front.

\begin{figure*}
  \includegraphics[width=0.5\textwidth]{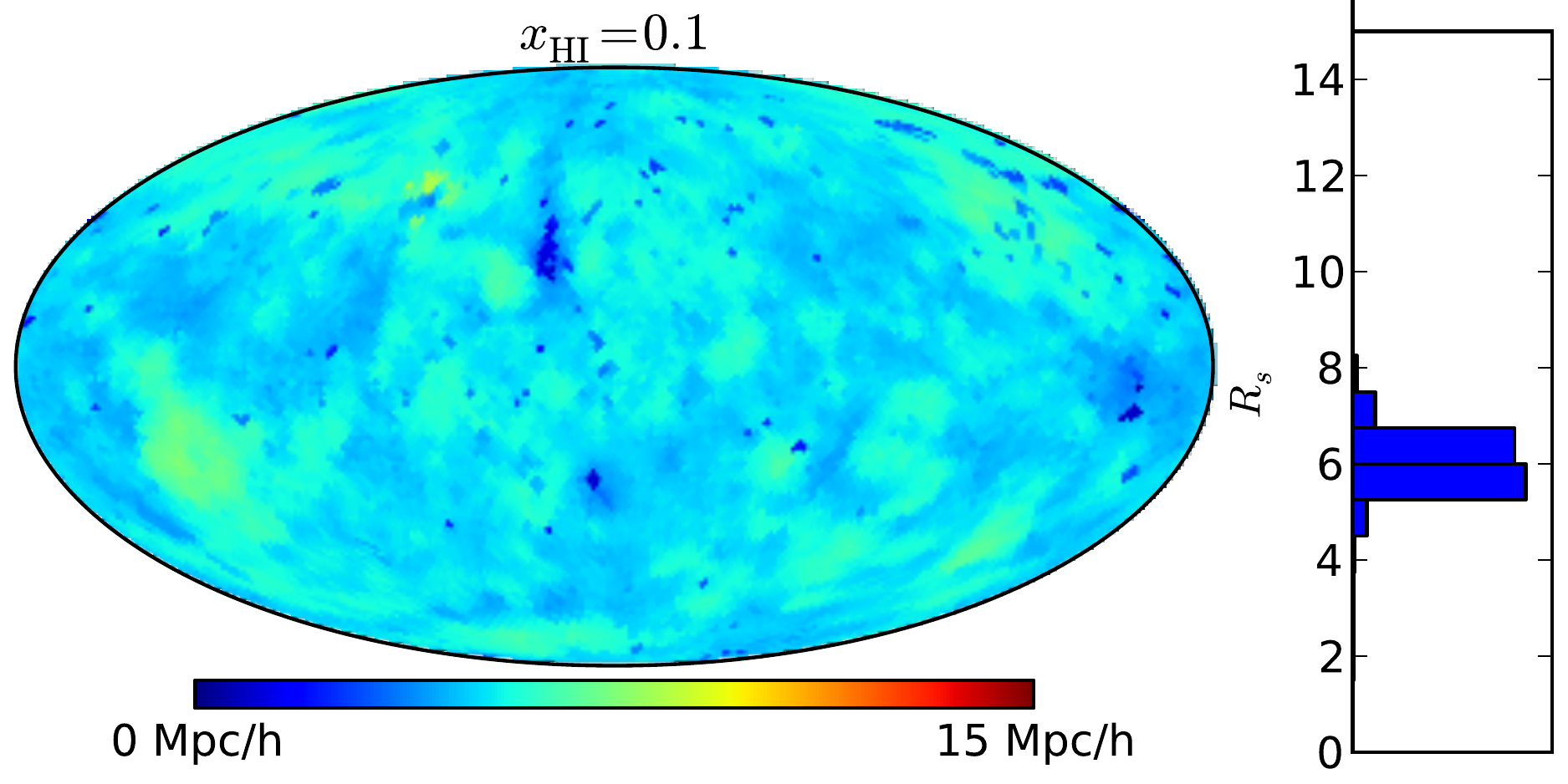}\includegraphics[width=0.5\textwidth]{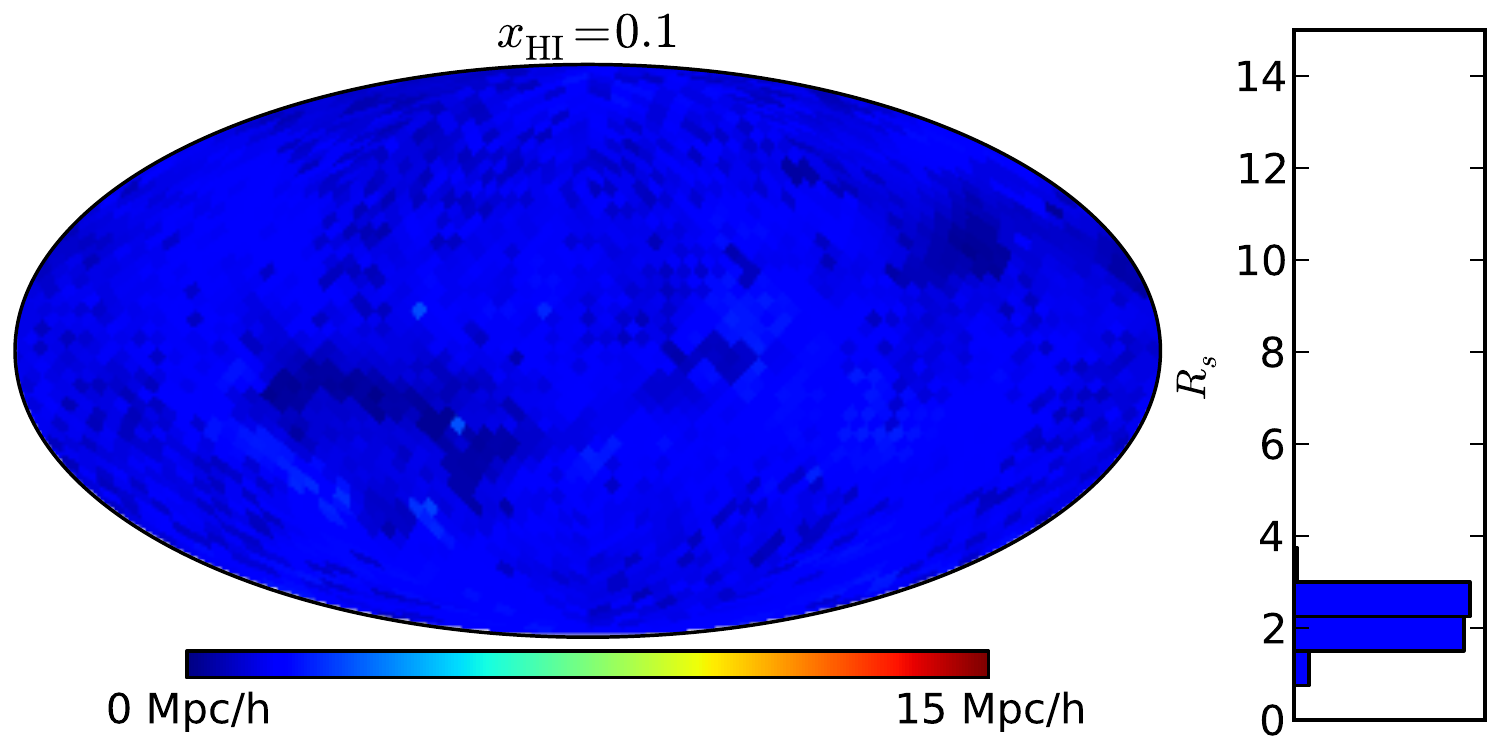}

  \includegraphics[width=0.5\textwidth]{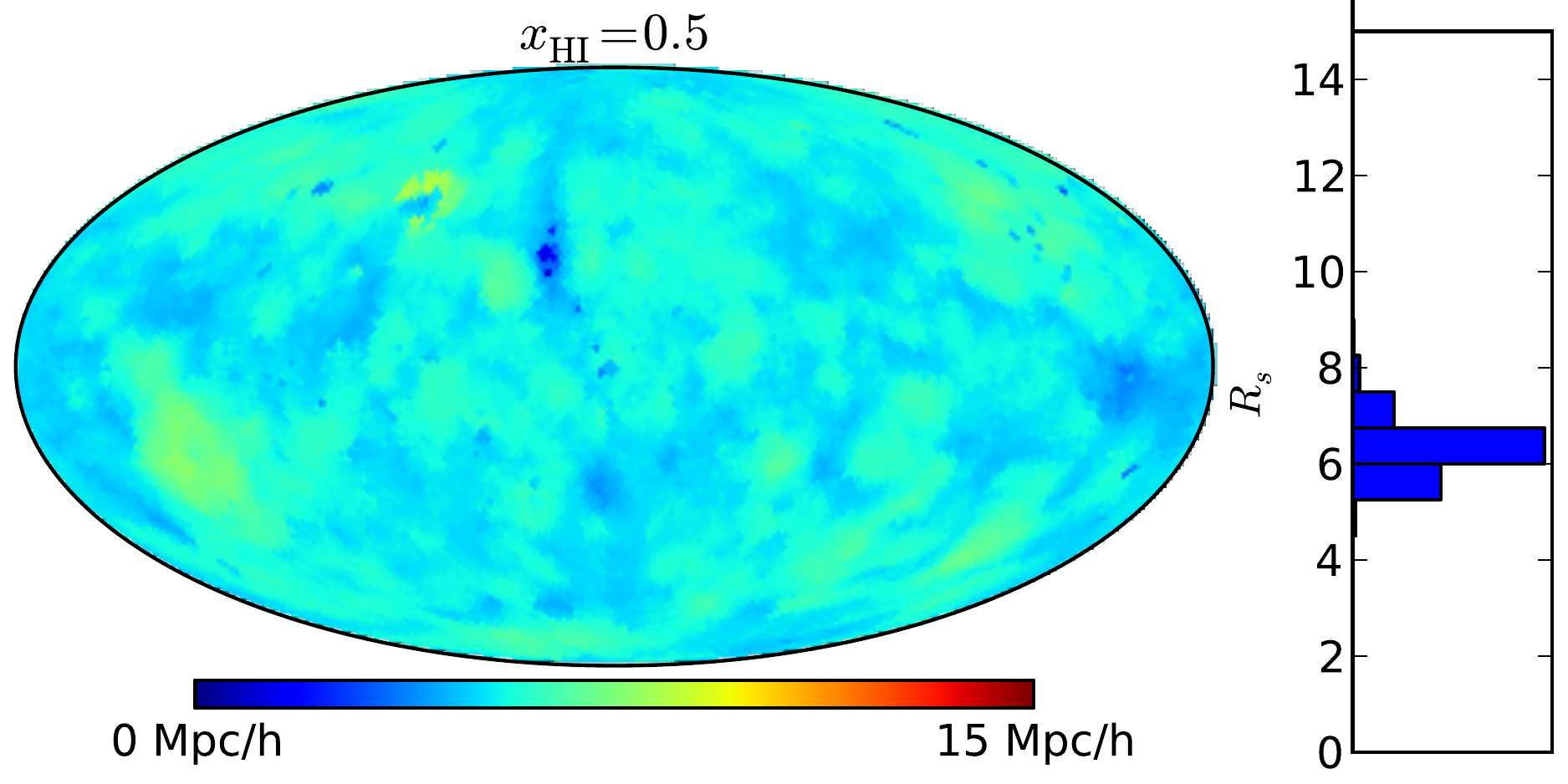}\includegraphics[width=0.5\textwidth]{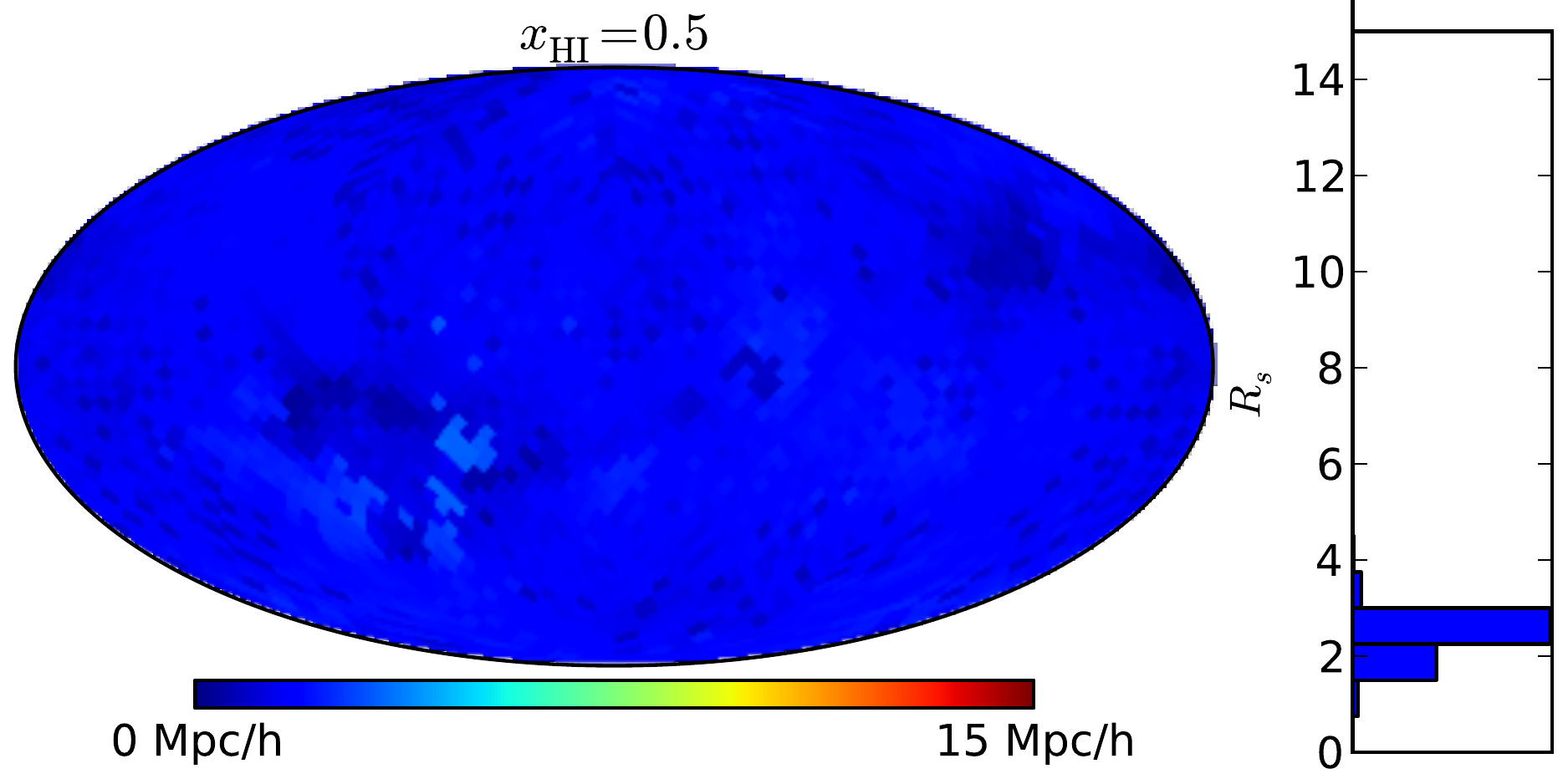}

  \includegraphics[width=0.5\textwidth]{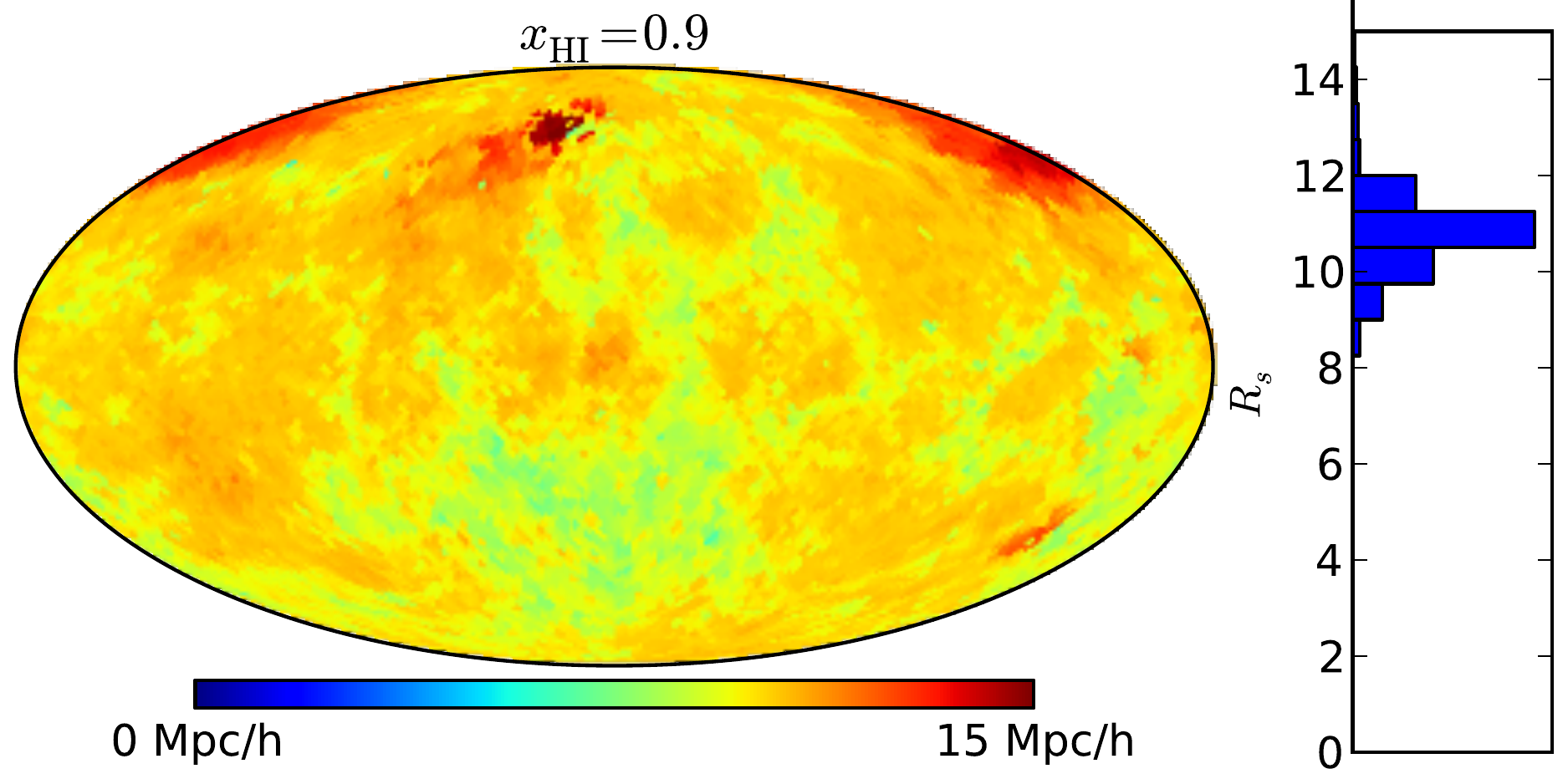}\includegraphics[width=0.5\textwidth]{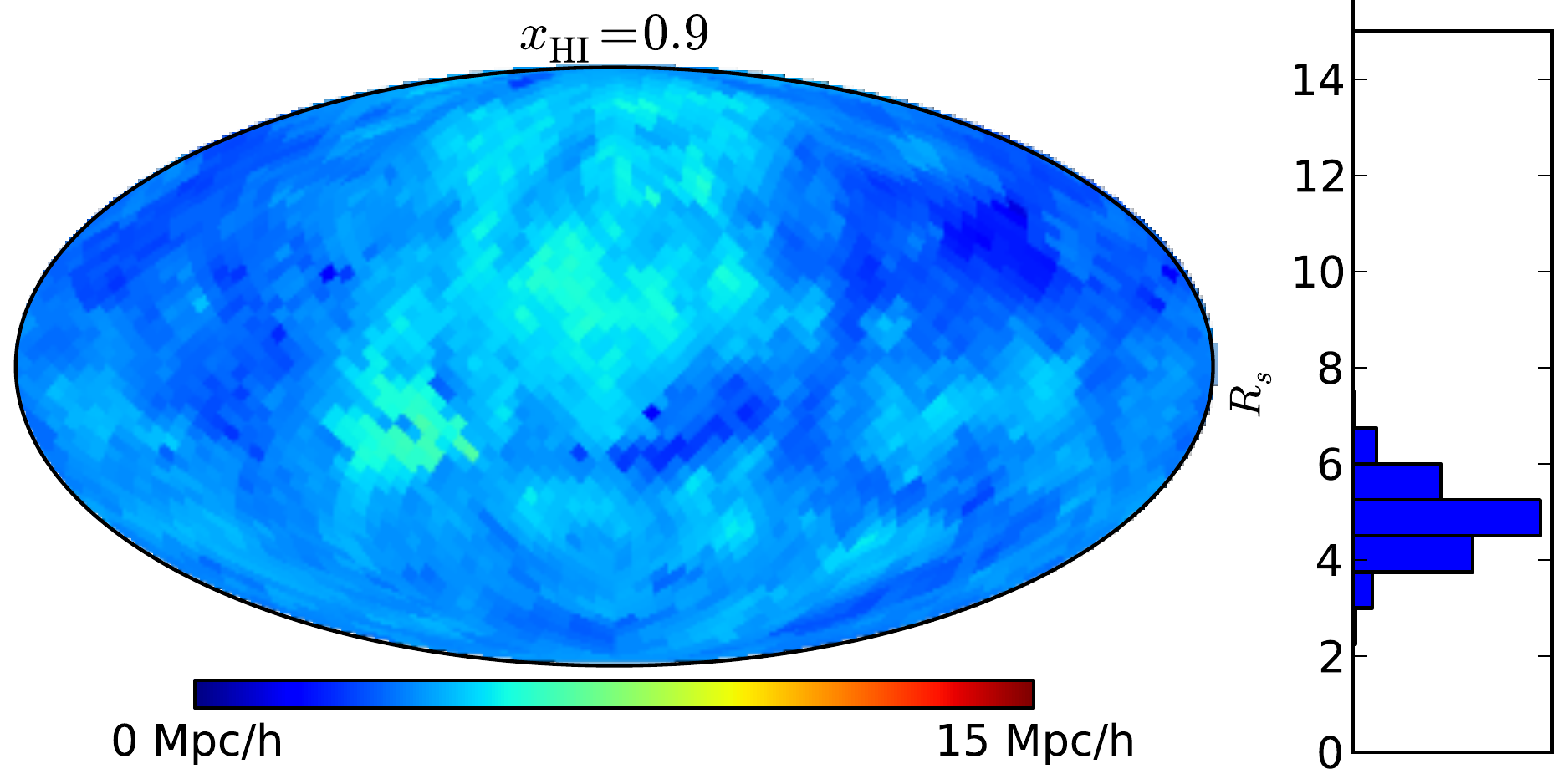}

  \includegraphics[width=0.5\textwidth]{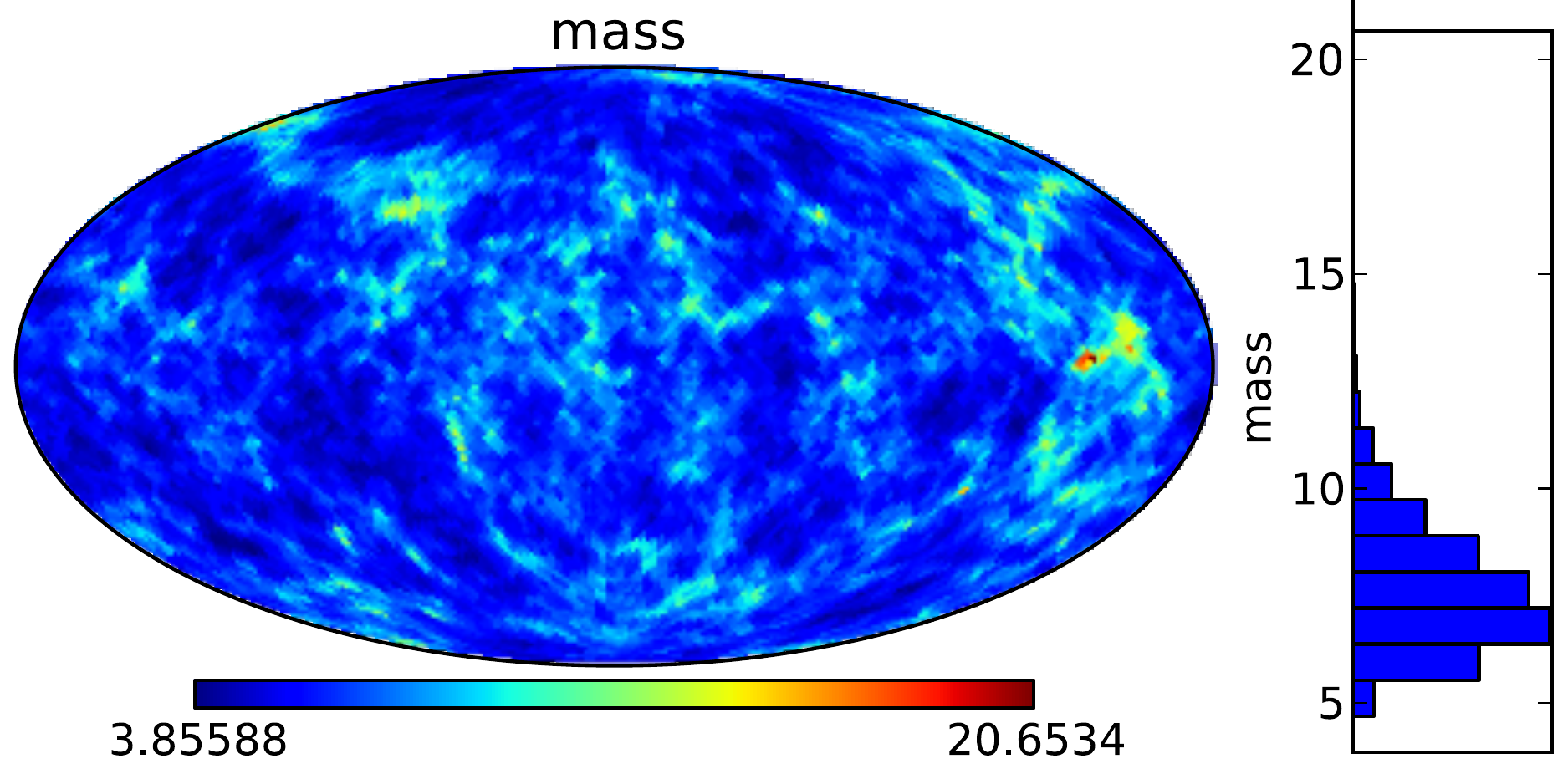}\includegraphics[width=0.5\textwidth]{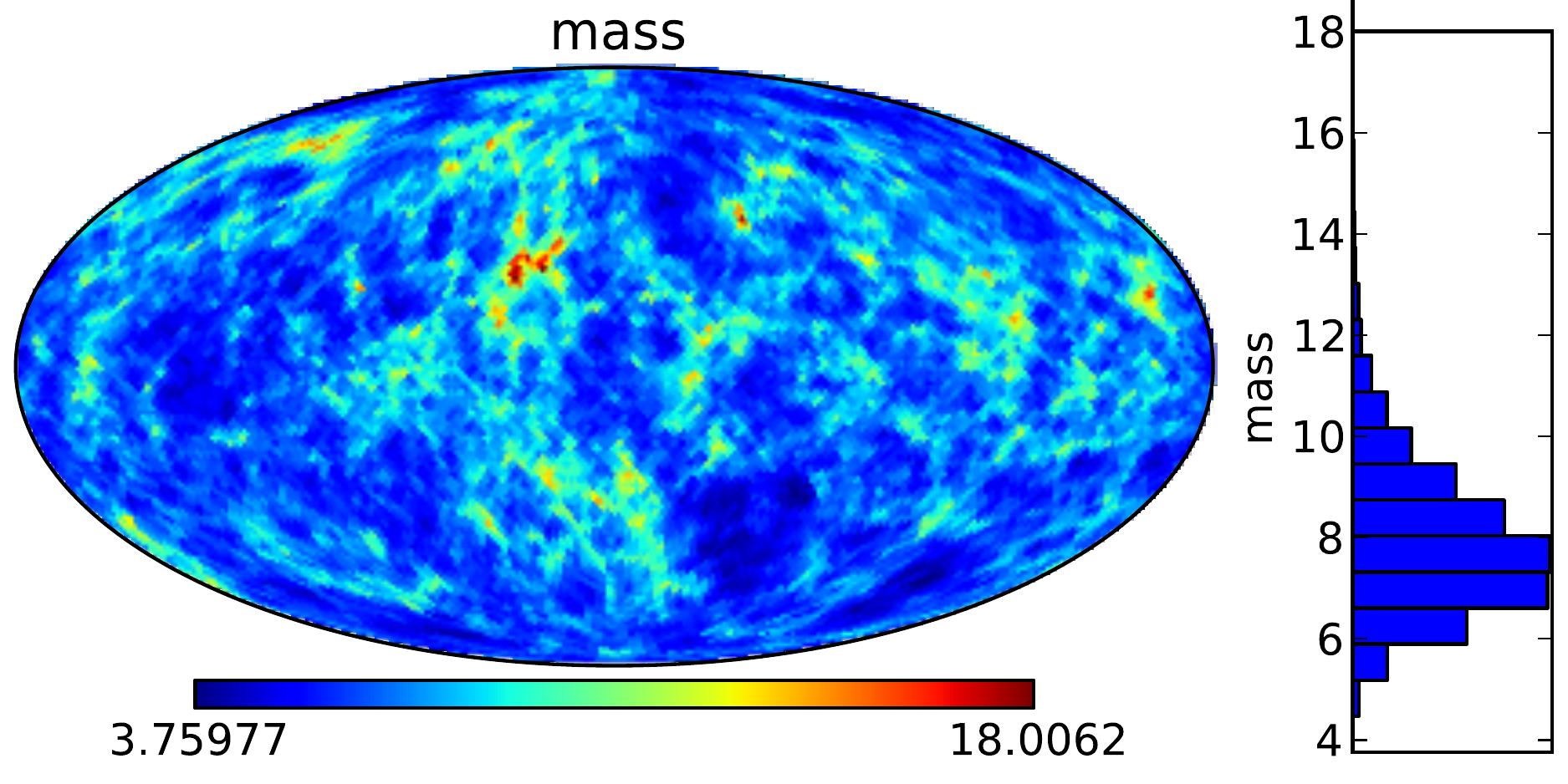}

  \caption{ Angular Distribution of the HII Radius.  The
  left column is sub-volume 0 (Q type) and the right column is
  sub-volume 4 (S type). From top to the bottom the inner
  ($x_{\rm HI=0.1}$), middle ($x_{\rm HI=0.5}$),
  and outer ($x_{\rm HI=0.9}$) fronts are shown. The
  histograms measure the variance of $R_{s}$ in different
  angles, characterizing the anisotropy. The red region in
  the $x_{\rm HI}=0.9$ plot of sub-volume 0 is due to
  merging with a further away HII bubble. The less
  significant red in the center corresponds to the merging
  shown in Figure \ref{fig:Evolution-of-if}.}
  \label{fig:Angular-Dependence-of}
\end{figure*}

In order to better visualize the structures which cause
    the anisotropy, we plot the distance to the different
    parts of the ionization fronts in a Molleweide projection
    as seen from the point of view of the central source.
    These maps of the angular-dependent HII bubble radius,
    $R_{s}$ are shown in Figure \ref{fig:Angular-Dependence-of},
    where we plot the quasar driven sub-volume 0 and stellar
    driven sub-volume 4.

Looking from the top panels downwards for sub-volume 0, we
    can first see clumps of mostly neutral gas close to the
    quasar which restrict the distance to the
    $x_{\rm HI}=0.1$ fronts to be very close by
    ($<\unit[1]{Mpc/h}$), compared to the mean distance for
    this neutral fraction of $~\unit[6]{Mpc/h}$. In the
    $x_{\rm HI}=0.9$ plot, for sub-volume 0 a prominent
    red region can be seen at the top. This corresponds to a
    bubble which has merged with the central bubble. Next to
    each panel in Figure \ref{fig:Angular-Dependence-of} we show a
    histogram of the $R_{s}$ values. The secondary bubble is
    responsible for the small tail of high values region in
    $x_{\rm HI}=0.9$ histogram, as well as giving an
    extra contribution to $A_{s}$.

Moving on to the stellar driven sub-volume 4 (right hand
    panels in Figure \ref{fig:Angular-Dependence-of}), we
    can see that the ionized region radius is smaller by
    approximately a factor of 3. Because this radius is
    small compared to the distance to nearby major halos,
    there is no sign of merging with other large bubbles,
    and the ionized region remains more isotropic.  In the
    bottom panels of Figure \ref{fig:Angular-Dependence-of}
    we show the Molleweide-projected mass density within the
    inscribed sphere of the sub-volume. It is interesting to
    compare this clumpiness with the structure in the
    ionized bubble radius. We can see some correlation with
    some structures, but not as much as might be expected,
    indicating that the interaction of radiation with the
    clumpy medium surrounding the sources is a complex
    process.

\begin{figure}
  \includegraphics[width=1\columnwidth]{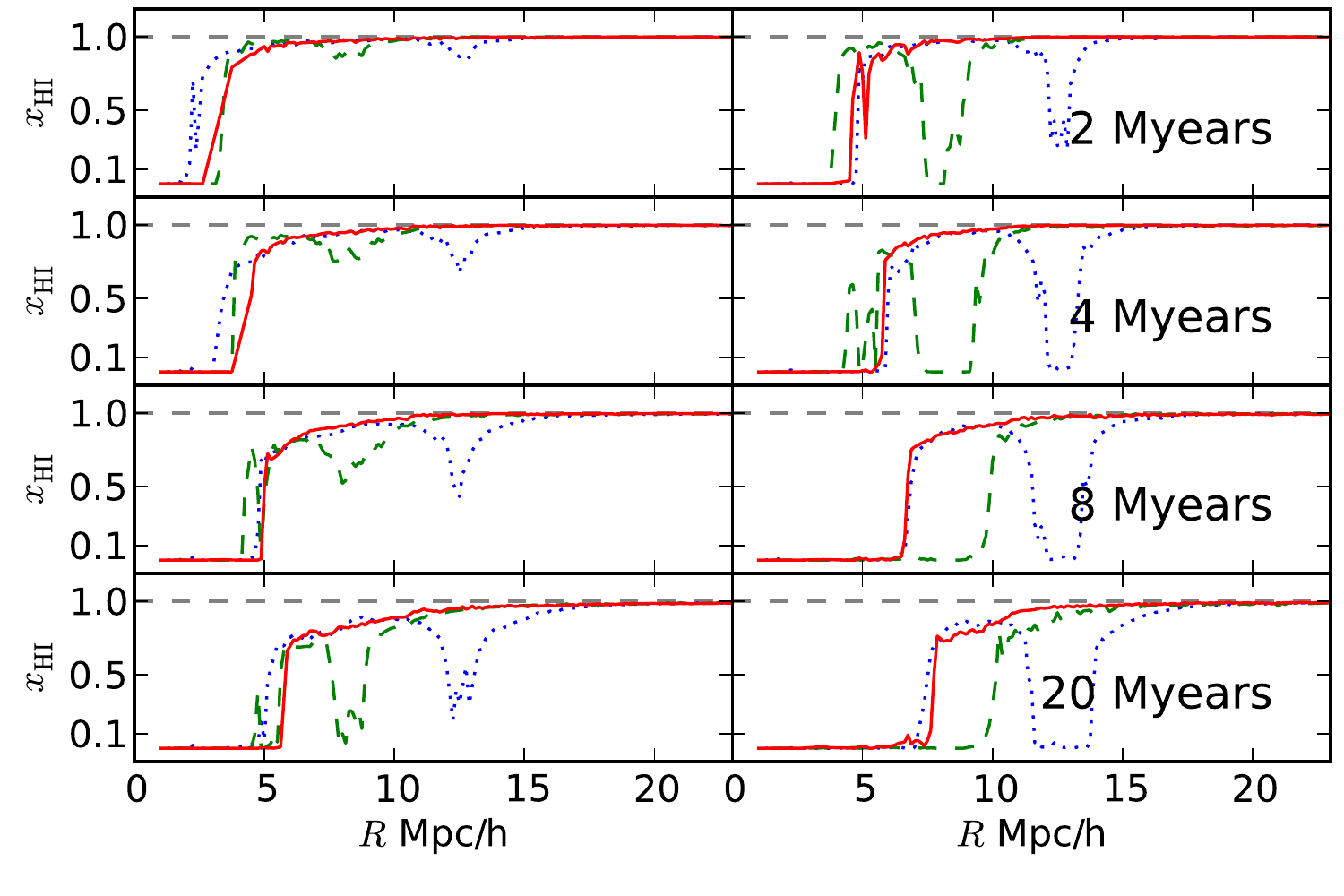}

  \caption{ Time evolution of $x_{\rm HI}$.  Time
  evolution of $x_{\rm HI}$ in sub-volume 0, along three
  selected sight-lines: (i) the blue line is directed towards
  the second brightest halo in the sub-volume; (ii) the green
  line is directed towards the merging bubble on the right
  shown in the slice plots; (iii) the red line is directed
  towards an arbitrary direction with no nearby bubbles.  The
  left panel is the luminosity model used throughout the
  paper, while on the right the stellar luminosity is
  boosted by a factor of 10 to demonstrate a stronger merging
  effect.}
  \label{fig:Time-evo-aniso}
\end{figure}

In Figure \ref{fig:Time-evo-aniso}, we show an orthogonal view to
    Figure \ref{fig:Angular-Dependence-of}, showing structure along
    rays moving out from the central source in some
    different directions. The outer HII front radii in three
    chosen directions are shown as a function of time. This
    allows us to see in detail how the clustering of halos
    affects the center bubble radius through merging as the
    simulation develops. The three directions are (i) one
    directed towards the second brightest source in the
    sub-volume, (ii) one directed towards the merging bubble
    seen on the right hand side of Figure \ref{fig:Evolution-of-if},
    and finally (iii) one oriented in an arbitrarily choosed
    direction not crossing any nearby secondary bubbles. In
    order to show the stronger bubble merging effect which
    would happen with more luminous sources, on the right
    hand side of Figure \ref{fig:Time-evo-aniso}, we plot
    the same 3 sight-lines, but after increasing the stellar
    luminosity by a factor of 10.

Merging happens as the central bubble touches the
    surrounding ones and meanwhile the radius along that
    direction significantly increases, compared to the
    radius without a nearby halo. This can be seen most
    clearly for sight-line (iii) on the right in
    Figure \ref{fig:Time-evo-aniso}.  The development of the
    anisotropy due to merging of nearby ionized bubbles is
    responsible for the apparent rapid growth of the
    averaged outer Stromgren radius after the growth of the
    inner radius is stabilized for S type and QS type
    sub-volumes. As the neutral fraction at the overlapping
    edge slowly drops below the threshold, two bubbles merge
    and the averaged radius increases. However, at the
    luminosity seen in the simulation (left hand panels of
    Figure \ref{fig:Time-evo-aniso}), such a merging
    mechanism is mostly limited to the growth of the outer
    front.  It is also worth noting that in the absence of
    the effect of smoothing due to secondary ionization the
    steep bubble edges will make the merging even more
    unlikely.

\section{Conclusions}
We have presented results from radiative transfer
    simulations in the vicinity of high redshift quasars in
    the MassiveBlack simulation.  We find that the rare
    brightest quasars drive a much more significant growth
    of ionized regions than in the purely stellar driven
    case. The ionized regions associated with active quasars
    are characterized by (i) a smooth ionized fraction
    transition from the middle to the outer front, and (ii)
    an increased anisotropy in the front when it starts to
    overlap the nearby ionized regions.  The nature of such
    growth is significantly more complex than a simple
    analytic growth of a single center bubble with clumping
    correction.

The largest HII bubble obtained in this simulation has a
    comoving radius of $\unit[10]{Mpc/h}$, which is smaller
    than the general expectation that can fulfill the
    reionization of the universe
    \citep{2011ASL.....4..228T, 2004Natur.432..194W,
    2011arXiv1111.6354M}. The quasar near zones we have
    presented in this paper are the primordial ancestors of
    the later much larger Stromgren spheres which will form
    near the end of the EoR. They are however relatively
    isolated regions that could be interesting objects for
    study in future 21cm surveys. After the $z=8$ epoch we
    have modeled in this paper, we expect that the global
    star formation in the MassiveBlack simulation increases
    significantly. This will eventually lead to the global
    reionization of the universe. We plan to study this
    process and role of quasars in future work.

\textbf{Acknowledgments:} We acknowledge support from Moore
    foundation which enabled the radiative transfer
    simulations to be run at the McWilliams Center of
    Cosmology at Carnegie Mellon University. The
    MassiveBlack simulation was run on the Cray XT5
    supercomputer Kraken at the National Institute for
    Computational Sciences. This research has been funded by
    the National Science Foundation (NSF) PetaApps program,
    OCI-0749212 and by NSF AST-1009781.  We also acknowledge
    the support of a Leverhulme Trust visiting professorship.
    The visualizations were produced using Healpix
    \citep{2005ApJ...622..759G}, Healpy
    \footnote{http://github.com/healpy/healpy}, and Gaepsi
    \footnote{https://github.com/rainwoodman/gaepsi}
    \citep{2011ApJS..197...18F}.  

\appendix{}
\section{Source Luminosity Models}
\label{sub:Luminosity-Model}
\subsection{Band luminosity of Quasars}
We calculate the bolometric luminosity $L_{{\rm bol}}$ of
    each quasar from the accretion rate $\dot{M}$ of the
    super-massive black hole with Equation
    \ref{eq:bh-luminosity}. We choose to use as $\dot{M}$ a
    single value throughout the calculation, which is
    averaged accretion rate measured from the simulation
    over a timescale $t_{Q}=\unit[2\times10^{7}]{yrs,}$
    where $t_{Q}$ is also the length of time over which we
    follow the evolution of the quasar ionization front. We
    estimate the flux in different bands from the bolometric
    luminosity according to the fitting formula of
    \citet{2007ApJ...654..731H}. The band definitions are
    listed in Table \ref{tab:Band-definitions.}.
\begin{table}
  \centering%
  \begin{tabular}{cc}
  \toprule 
  Band & Range\tabularnewline
  \midrule
  \midrule 
  UV & $\unit[13.6]{eV}$ to $\unit[250]{eV}$\tabularnewline
  \midrule
  \midrule 
  Soft X-Ray & $\unit[250]{eV}$ to $\unit[2]{KeV}$\tabularnewline
  \midrule
  \midrule 
  Hard X-Ray & $\unit[2]{eV}$ to $\unit[10]{KeV}$\tabularnewline
  \bottomrule
  \end{tabular}
  
  \caption{\label{tab:Band-definitions.}Band definitions.}
\end{table}
The soft and hard X-Ray band luminosity are directly
    calculated with the program distributed by
    \citet{2007ApJ...654..731H}.  We obtain the UV band
    luminosity from the reported B band luminosity following
    the broken power law described in the paper
    ($\alpha_{\rm UV}=1.76$ and
    $\alpha_{\rm B}=0.44$).
\begin{equation}
  L_{\rm UV} = 
   \frac{L_{B}}{\nu_{B}}
   \left(\frac{\nu_{X}}{\nu_{B}}\right)^{-\alpha_{B}}
   \left(\frac{\nu_{I}}{\nu_{X}}\right)^
         {-\alpha_{\rm UV}}
   \frac{\left(\frac{250}{13.6}\right)^{1-\alpha_{\rm UV}}-1}
   {1-\alpha_{\rm UV}},
\end{equation}
    where $\nu_{X}=\unit[c/120]{nm}$ is the pivot from
    optical to UV, $\nu_{B}=\unit[c/445]{nm}$ and
    $\nu_{I}=\unit[c/91.1]{nm}$. We calculate the photon
    flux of the quasar sources according to these power law
    spectra applied to the various bands.

\subsection{Flux of UV stellar photons \label{sub:Stellar-Luminosity}}

The stellar photon flux of a halo is given by 
\begin{equation}
  \dot{N}_{\rm SFR}=f_{\rm esc}N_{\gamma/\rm H}X_{\rm H}\Phi/m_{p},
\end{equation}
    where $\Phi$ is the star formation rate of the halo,
    $X_{\rm H}=0.76$, and $m_{p}$ is the proton mass. We take
    the escape fraction to be $f_{\rm esc}=0.1$, and
    the photon to hydrogen ratio to be $N_{\gamma/\rm
    H}=4000$, a value found to produce a reionization
    history consistent with observations
    \citep{2004ApJ...613....1F, 2003MNRAS.344..607S}. For
    simplicity, the stellar sources are assigned to the
    center of the halo; in future work we plan to
    investigate the effect of the positioning of the stellar
    sources by either following the star forming gas
    particles in the simulation directly or the
    gravitationally bound sub-halos or galaxies. To ease the
    modeling, we only include and assume the same UV band
    spectrum for the stellar radiation to that of the quasar
    spectrum.

\bibliographystyle{mn2e}
\bibliography{mybib}
\end{document}